\documentclass[]{emulateapj}
\usepackage[backref, breaklinks, colorlinks, citecolor=blue, linkcolor=magenta]{hyperref}
\usepackage[dvipsnames]{xcolor}

\slugcomment{Submitted to The Astrophysical Journal}

\shorttitle{BH Driven \ion{O}{6} in the CGM}
\shortauthors{Sanchez et. al.}

\usepackage{natbib}
\begin{document} 

\title{Not So Heavy Metals: Black Hole Feedback Enriches The Circumgalactic Medium}

\author{N. Nicole Sanchez\altaffilmark{1}}
\author{Jessica K. Werk\altaffilmark{1}}
\author{Michael Tremmel\altaffilmark{2}}
\author{Andrew Pontzen\altaffilmark{3}}
\author{Charlotte Christensen\altaffilmark{4}}
\author{Thomas Quinn\altaffilmark{1}}
\author{Akaxia Cruz\altaffilmark{1}}

\affil{$^1$Astronomy Department, University of Washington, Seattle, WA 98195, US, sanchenn@uw.edu}
\affil{$^2$Yale Center for Astronomy \& Astrophysics, Physics Department, P.O. Box 208120, New Haven, CT 06520, USA}
\affil{$^3$Department of Physics \& Astronomy, University College London, 132 Hampstead Road, London, NWI 2PS, United Kingdom}
\affil{$^4$Physics Department, Grinnell College, 1116 Eighth Ave., Grinnell, IA 50112, United States}

% #####################################
% ############# ABSTRACT ##############
% #####################################
\begin{abstract}\label{abs:abstractlabel}
 
We examine the effects of supermassive black hole (SMBH) feedback on the circumgalactic medium (CGM) using a cosmological hydrodynamic simulation \citep[{\sc Romulus25};][]{Tremmel2017} and a set of four zoom-in ``genetically modified'' Milky Way-mass galaxies sampling different evolutionary paths. By tracing the distribution of metals in the CGM, we show that \ion{O}{6} is a sensitive indicator of SMBH feedback. First, we calculate the column densities of \ion{O}{6} in simulated Milky Way-mass galaxies and compare them with observations from the COS-Halos Survey. Our simulations show column densities of \ion{O}{6} in the CGM consistent with those of COS-Halos star forming and quenched galaxies. These results contrast with those from previous simulation studies which typically underproduce CGM column densities of \ion{O}{6}. We determine that a galaxy's star formation history and assembly record have little effect on the amount of \ion{O}{6} in its CGM. Instead, column densities of \ion{O}{6} are closely tied to galaxy halo mass and black hole growth history. The set of zoom-in, genetically modified Milky Way-mass galaxies indicates that the SMBH drives highly metal-enriched material out into its host galaxy's halo which in turn elevates the column densities of \ion{O}{6} in the CGM. 

\end{abstract}

% ########## END OF ABSTRACT ##########
% #####################################

% #####################################
% ########## INTRODUCTION #############
% #####################################
\section{Introduction} 
\label{sec-intro}

The circumgalactic medium (CGM), the extended region of gas surrounding galaxies out to their virial radii, is richly structured and composed of the raw materials and by-products of galaxy evolution. Due to its extremely diffuse nature, the CGM is the component of a galaxy that presents perhaps the greatest challenge to extragalactic observers. The most sensitive probes of the predominantly ionized gas in the CGM are background QSO sightlines. The spectra of these background QSOs show the absorption signature of a foreground galaxy's halo \citep[e.g.][]{Bahcall1969, Bergeron1986}. Such studies provide an inherently one-dimensional picture of the gas, typically along only a single sightline.  Other observational techniques of studying the CGM include: stacking analyses, which combine between hundreds and thousands of spectra and/or images to detect the faint signals of CGM \citep{York2006,Peek2015,Steidel2010, Zhu2013, Zhang2018}; ``down-the-barrel'' spectroscopy, which employs a galaxy's own starlight as the background source for CGM absorption \citep{Martin2006,Bordoloi2011,Rubin2014,Heckman2015}; and emission line maps, which search for the few photons emitted directly by CGM gas \citep{Putman2012a,Cantalupo2014,Hayes2016}. Additionally X-ray observations by Chandra and XMM Newton have been used to help constrain the extent and nature of the hot, 10$^{6}$ K CGM \citep[e.g.][]{Nicastro2005, Anderson2010, Yao2010, Anderson2013}. 

Significant progress in the study of the $z$ $\lesssim$ 1 CGM followed the 2009 installation of the UV-sensitive Cosmic Origins Spectrograph (COS) on the \textit{Hubble Space Telescope} \citep[{\emph{HST},}][]{Green2012}. Numerous successful absorption-line surveys with COS have reported a structurally complex, multiphase medium with column densities and covering fractions of metal ions and hydrogen depending strongly on galaxy properties \citep[e.g.][]{Tripp2011, Stocke2013, Tumlinson2013, Borthakur2015, Liang2014, Johnson2015, Bordoloi2014, Keeney2017, Prochaska2017}.  For example, while actively star-forming galaxies exhibit a highly-ionized component to their CGM characterized by strong \ion{O}{6} absorption out to at least 150 kpc, non-star-forming, elliptical galaxies show weak or no detections of \ion{O}{6} \citep{Tumlinson2011}. However, these same passive galaxies exhibit a high incidence of strong \ion{H}{1} absorption in their CGM, as much cold, bound gas as their star-forming counterparts \citep{Thom2012, Prochaska2013, Johnson2015}.  These results emphasize that the processes that transform galaxies from star-forming disks to passive ellipticals {\emph{do}} affect the physical state of the CGM, but {\emph{do not}} completely deplete it of cool, 10$^{4}$ K gas. 

Numerous studies of the CGM indicate that it hosts a substantial fraction of a galaxy's baryons \citep[e.g.][]{Werk2014, Keeney2017, Prochaska2017}. Overall, the observational studies on the low-redshift CGM all highlight the driving role played by gas in the galactic halo in shaping the evolution of stars and gas in the disks. It is clear that understanding the CGM is crucial for understanding the complex nature of galaxy evolution and growth. 

The widespread \ion{O}{6} absorption in MW-mass halos, referenced above, has presented a particularly intriguing puzzle for theorists \citep[e.g.][]{Stern2016, Suresh2017, Oppenheimer2016, Mcquinn2018, Nelson2018}.  \cite{Oppenheimer2016} argue that the \ion{O}{6} bimodality in SF vs non-SF halos  arises due to collisionally-ionized \ion{O}{6} acting as tracer of the virial temperature of gas in these galaxy halos. In this scenario, galaxies with M$_{*}$ $\gtrsim$ 10$^{11}$ M$_{\odot}$ would have more of their oxygen in a more ionized phase such as \ion{O}{7} and \ion{O}{8}. This hypothesis is supported by observations of non-SF galaxies in the COS-Halos sample, which show lower column densities of OVI, reportedly due to the intrinsically higher virial temperatures of these generally more massive, red ellipticals. In contrast, \cite{Suresh2017} argue that \ion{O}{6} is built up by supermassive black holes (SMBH), which can physically modify the CGM via outflows and heat it to 10$^{5.5 - 5.8}$ K, the temperature at which the fraction of oxygen as \ion{O}{6} is maximized. Meanwhile, \cite{Oppenheimer2018} suggests that photo-ionizing energy from a flickering AGN might be required to raise the column densities of OVI within the virial radius to observed levels.  These two pictures differ greatly in terms of the physical processes that give rise to widespread \ion{O}{6} in the CGM. In one, \ion{O}{6} traces the hot halo that forms in conjunction with the galaxy itself though gravitational processes. In the other, pc-scale processes in the inner, central galaxy provide enough heat and energy to impact the physical state of gas in its extended halo out to 100 kpc.

With respect to the picture put forward by \cite{Suresh2017}, it is not unreasonable to propose that a galaxy's SMBH influences the content of its CGM.  Galaxy properties in general have been shown to be strongly tied to the evolution of its central SMBH. Relations such as the M-$\sigma$ and the bulge mass-BH mass correlation \citep{Ferrarese2000,Mcconnell2013} indicate that the SMBH and its host galaxy halo \textit{co-evolve} \citep[][and references therein]{Gebhardt2000,Volonteri2012b,Kormendy2013,Reines2015c}.  However, the direct mechanisms for SMBH-CGM impact remain unclear. 

SMBHs have been proposed to effect the CGM in a variety of ways. First, feedback from the active SMBH may inject energy into the surrounding material, raising temperatures, resulting in collisionally- and photo-ionized metals in the gas \citep{McQuinn2017,Mathews2017,Oppenheimer2018}. Additionally, the SMBH may physically push multiphase gas out of the galaxy. Some of this material may end up falling back into the galaxy as part of the ``recycling'' of the CGM \citep{Tumlinson2017}, enriching CGM gas with metals from the center of the galaxy, or may leave the CGM entirely and instead enrich the intergalactic medium (IGM). 

In tandem with observational progress on characterizing the CGM, cosmological hydrodynamic simulations have become a powerful tool for examining the physics driving the multiphase nature of the CGM \citep{Stinson2012, Shen2012, Hummels2013, Cen2013, Ford2014, Ford2016b, Oppenheimer2016, Liang2018, Suresh2017,Nelson2018}. Despite significant effort, few of these studies are able to match the observed properties of the CGM.  For example, most previous studies underpredict the column densities of \ion{O}{6} found by COS-Halos  \citep[including the aforementioned studies,][]{Oppenheimer2008, Hummels2013, Oppenheimer2016, Suresh2017}. Nonetheless, these studies have led to important physical insights. Using the smooth particle hydrodynamic code GADGET-2 \citep{Springel2005a,Oppenheimer2008}, \cite{Ford2014} found that the presence of \ion{O}{6} in the CGM likely arises from metals ejected very early on in the galaxy's evolution.  More recently, \cite{Nelson2018} matched the COS-Halos observations using the IllustrisTNG simulations and determined that the amount of \ion{O}{6} the CGM can depend on a variety of galactic properties including sSFR. In particular, they find that BH feedback (specifically, their low-accretion, kinetic-feedback mode) plays a crucial role in setting the amount of \ion{O}{6} in the CGM by affecting the amount of metal mass ejected by the galaxy.

Motivated by previous theoretical and observational work, we use two sets of simulations to study circumgalactic \ion{O}{6}: the cosmological volume, Romulus25 \citep{Tremmel2017}, and three ``genetically modified'' variations of a zoom-in Milky-Way (MW) mass galaxy \citep{Roth2016,Pontzen2017} selected from a cosmological volume. These genetically modified zoom-in galaxies are run with and without the implementation of BH physics to test the effect of SMBH feedback on the CGM. To compare our results with observations, we rely primarily on data from the COS-Halos Survey. Although several other surveys have examined the CGM around a wide-range of galaxies \citep[e.g.][]{Stocke2013, Savage2014, Borthakur2015, Danforth2016, Keeney2017}, COS-Halos \citep{Tumlinson2013} remains the best-studied, uniformly-selected sample of MW-mass host galaxies to-date, and one of the few to focus on \ion{O}{6}. Furthermore, COS-Halos tabulates CGM gas column densities along with spectroscopically and photometrically-determined galaxy properties (e.g. SFR, M$_{*}$) allowing for a straightforward comparison between our simulations and the data. 

Ultimately, we examine the effects of both environmental and internal galaxy processes on the physical state and content of the CGM. Specifically, we address how the star formation and assembly history of the galaxy impact the content of the CGM and how SMBH activity imprints itself on the CGM. Using these zoom-in simulations in tandem with the {\sc Romulus25}  simulation, we illuminate the roles that stellar evolution and SMBH feedback play in setting the properties of the CGM of MW-mass galaxies.

In Section \ref{sec-model}, we describe the underlying physics used in our two galaxy samples. Section \ref{sec-results} details our results from examining the CGM in ROMULUS25 and comparisons with the zoom-in galaxies and observations. We discuss these results and their implications for future studies in Section \ref{sec-discuss}. In Section \ref{sec-conclude}, we summarize and offer conclusions. 

% ####### END OF INTRODUCTION #########
% #####################################

% #####################################
% ####### SIMULATION PARAMETERS #######
% #####################################
\section{Simulation Parameters}
\label{sec-model}

\subsection{ChaNGa Physics} 
Both {\sc Romulus25}  (hereafter R25) and our set of zoom-in galaxies were run using the smoothed particle hydrodynamics N-body tree code, Charm N-body GrAvity solver \citep[ChaNGa,][]{Menon2015}. ChaNGa includes the same models for a cosmic UV background, star formation (using a Kroup IMF), `blastwave' SN feedback, and low temperature metal line cooling as previously used in GASOLINE \citep{Wadsley2004,Wadsley2008,Stinson2006,Shen2010}. Neither {\sc Romulus25}  or the zoom-in simulations utilize metal cooling as the resolution of these simulations is too large to consider individual star forming regions. Instead, our simulations use a low temperature extension to the cooling curve such that only gas below 10$^{4}$ K cools proportionally to the metals in the gas. Gas above this threshold cools only via \ion{H}{0}/\ion{He}{0}, Bremsstrahlung, and inverse Compton. This lack of metal cooling in our model likely causes our galaxies to over predict \ion{O}{6}  by approximately 0.3 dex \citep{Shen2010}; however, as our study compares total quantities of oxygen between simulations and the relative motions of metals, the relative values of N$_{\rm OVI}$ between simulations remains valid.

ChaNGa includes an improved SPH formalism which includes a geometric density approach in the force expression \citep{Wadsley2017}. This update to the hydrodynamic treatment includes thermal diffusion \citep{Shen2010} and reduces artificial surface tension allowing for better resolution of fluid instabilities \citep{Ritchie2001,Menon2015,Governato2015}. 

Additional improvements have been made to the BH formation, accretion, and feedback models as well an improved prescription for dynamical friction \citep{Tremmel2015,Tremmel2017}. BH seed formation is tied to dense, extremely low metallicity gas to better estimate SMBH populations in a wide range of galaxies. Sub-grid models for both dynamical friction\textemdash to better simulate realistic SMBH dynamical evolution and mergers\textemdash and accretion have been implemented. The new SMBH accretion model considers angular momentum support from nearby gas allowing for more physical growth compared to Bondi-Hoyle prescription alone or other methods that require additional assumptions or free parameters \citep{Rosas-Guevara2015,Angles-Alcazar2017}. Angular momentum support is taken into account in the accretion equation:
\begin{equation}
\dot{M} \propto \frac{\pi (GM)^2 \rho c_s}{(v_{\theta}^2 + c_s^2)^2},
\end{equation}
where $v_{\theta}$ is the rotational velocity of the gas surrounding the BH and is informed by the angular momentum support of the gas on the smallest, resolvable scale. However, when bulk motion dominates over rotational motion, the formula reverts to the original Bondi-Hoyle.
Thermal SMBH feedback energy is imparted on the nearest 32 gas particles according to a kernel smoothing and is determined by the accreted mass, $\dot{M}$, as: 
\begin{equation}
E = \epsilon_{r} \epsilon_{f} \dot{M} c^2 dt, 
\end{equation}
where $\epsilon_{r}$ = 0.1 and $\epsilon_{f}$ = 0.02 are the radiative and feedback efficiency, respectively, and $dt$ represents one black hole (BH) timestep, during which the accretion is assumed to be constant. Cooling is briefly ($\sim$ 10$^{4-5}$ years) shut off immediately after AGN feedback events. \citep{Tremmel2017} Our SMBH feedback prescription is also shown to be able to produce large scale outflows \citep{Pontzen2017,Tremmel2018}. 

All our simulations were run with a $\Lambda$CDM cosmology from the most recent Planck collaboration utilizing $\Omega_0$ = 0.3086, $\Lambda$ = 0.6914, h = 0.6777, $\sigma_8$ = 0.8288 and have Plummer equivalent force softening lengths of 250 pc. For simulating the cosmic reionization energy, both simulations have a \cite{Haardt2012} UV background applied at $z$ $\sim$ 9 through the evolution to low-redshift. For our purposes, we've defined the CGM in each simulated galaxy as all the gas inside the galaxy's virial radius, defined as the radius at which the density is 200 times the critical density, $\rho_{c}$, where $\rho/\rho_{c}$ = 200 (R$_{200}$), and outside a spherical 10 kpc from its center.

\subsection{{\sc Romulus25}  Cosmological Volume}
\begin{figure}[]
\centerline{\resizebox{1.02\hsize}{!}{\includegraphics[angle=0]{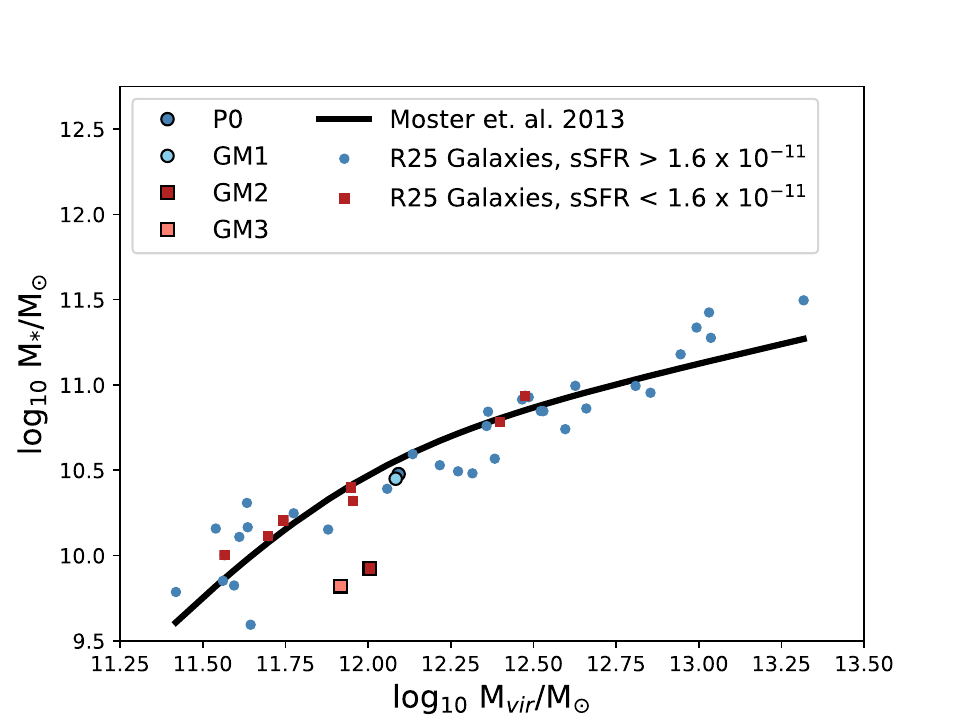}}}
\caption[]{We show that the 39 galaxies from R25 in our sample, which are selected along the distribution of COS-Halos stellar masses within the range (3 $\times$ 10$^{9}$ M$_{\odot}$ \textemdash 3 $\times$ 10$^{11}$ M$_{\odot}$). Including the corrections of \cite{Munshi2013}, the galaxies follow the stellar mass-halo mass (SHMH) relation up to $\sim$ 10$^{13}$ above which they are slightly higher than predicted. Red squares and blue circles represent passive and star forming galaxies, respectively. The 4 zoom-in galaxies with BH physics are outlined in black.}
\label{fig-massmetal}
\end{figure}

The {\sc Romulus25}  \citep[][R25]{Tremmel2017} simulation is a 25 Mpc cosmological volume which includes galaxy halos within the mass range 10$^{9}$ \textemdash 10$^{13}$ M$_{\odot}$. R25 has a mass resolution of 3.4 $\times$ 10$^5$ M$_{\odot}$ and 2.1 $\times$ $10^5$ M$_{\odot}$ for DM and gas particles, respectively. Galaxies in R25 have been shown to lie along the M$_{BH}$-M$_{*}$ and stellar mass-halo mass relation (Figure \ref{fig-massmetal}, though slightly higher than predicted for the highest mass galaxies),  and are consistent with observations of star formation and SMBH accretion histories at high redshift \citep{Tremmel2017}. Both our M$_{halo}$ and M$_{*}$ measurements use the corrections from \cite{Munshi2013}. Additionally, \cite{Tremmel2017} shows that SMBH physics is a necessary component for reproducing the evolution of MW-mass galaxies as well as quenching in massive galaxies. For our study, we focus galaxies in R25 that fall within the stellar mass range of COS-Halos: 3 $\times$ 10$^{9}$ M$_{\odot}$ and 3 $\times$ $10^{11}$ M$_{\odot}$ and populate a similar distribution of stellar masses. 

With these selection criteria in place, our sample includes 39 galaxies. Using the specific star formation (sSFR = SFR/M$_{*}$) cut of COS-Halos, 32 of these galaxies are star forming (sSFR \textgreater \ 1.64 $\times$ 10$^{-11}$ yr$^{-1}$) and 7 are passive at $z$ $\sim$ 0.17. The sSFR of COS-Halos (previously \textgreater  10$^{-11}$ yr$^{-1}$) has been corrected by a factor of 1.64 to account for the fact that COS-Halos uses a Salpeter IMF while our simulations use a Kroupa IMF \citep{Kroupa2001a}. This correction only affects the categoration two of our R25 galaxies. We further note that this fraction of passive galaxies is a conservative estimate. However, by $z$ = 0, the quenched fraction in R25 is about 40 \% for the highest mass galaxies \citep{Tremmel2018b}.

\begin{figure}[]
\vspace{-2mm}
\centerline{\resizebox{1.02\hsize}{!}{\includegraphics[angle=0]{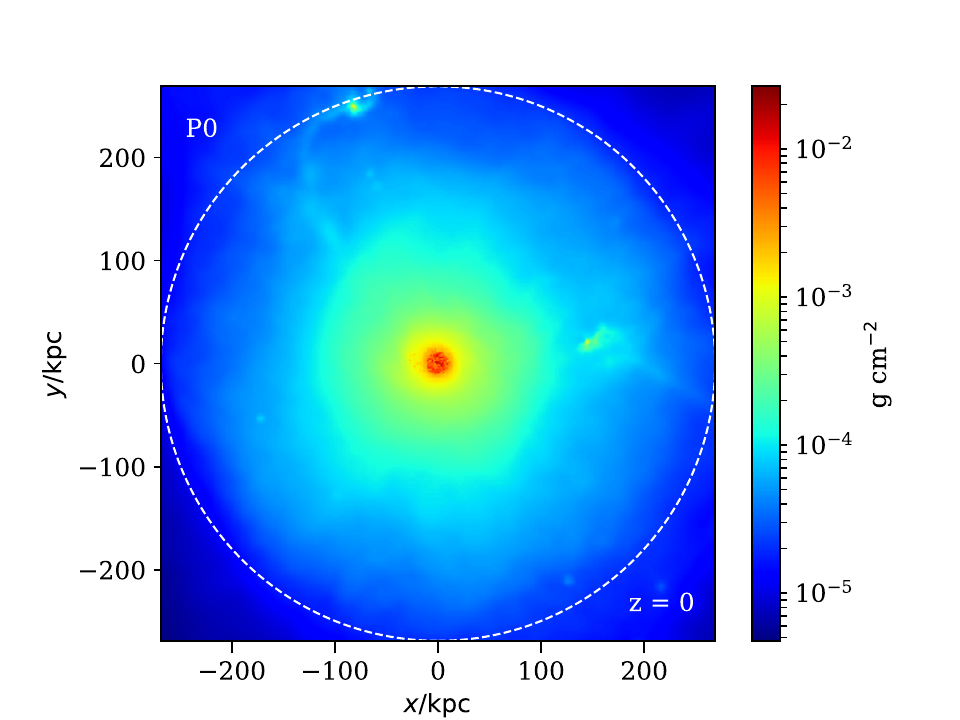}}}
\vspace{0mm}
\centerline{\resizebox{1.02\hsize}{!}{\includegraphics[angle=0]{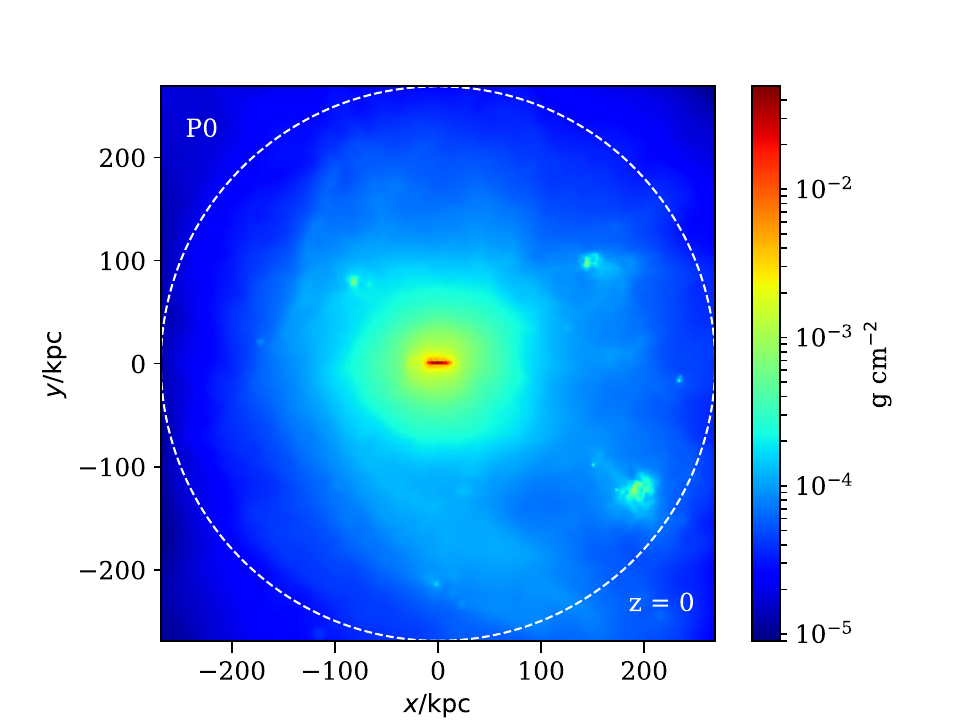}}}
\caption[]{A face-on and edge-on view of our Patient 0 galaxy in projected gas density at $z$ = 0.. The virial radius is designated by the white-dashed circle.}
\label{fig-P0}
\end{figure}

\subsection{Zoom-In Galaxies: Patient 0 and its Genetic Modifications}
While R25 gives a cosmological context to our analysis, we examine our set of genetically modified zoom-in galaxies to better understand the physical and phenomenological processes that influence the CGM. To select our MW-mass galaxy, we ran an initial dark-matter-only simulation in a 50-Mpc-on-a-side cosmological volume. From this simulation, we selected a MW-analog (M$_{vir}$ = 9.9 $\times$ 10$^{11}$ M$_{\odot}$) halo at z=0 as our ``Patient 0'' (hereafter P0) and then re-simulated it at a higher resolution with baryons. We additionally required that the galaxy be \textgreater 2 Mpc away from another MW- or higher mass galaxy. Finally, we selected our P0 for the satellite galaxy (M$_{sat}$ = 2 $\times$ 10$^{10}$ M$_{\odot}$) contained within its virial radius at $z$ = 0 which acts as a proxy for a LMC satellite. Selecting a MW-analog galaxy in halo mass allows us to compare directly with the COS-Halos observations, which observed $\sim$ L$^{*}$ galaxies. For the subsequent, ``genetically modified'' (GM) zoom-in runs, we use the method of genetic modification of \cite{Pontzen2016} which creates a set of very similar initial conditions that result in galaxy simulations which keep the large scale structure and cosmological conditions consistent (as in P0), while resulting in slight modifications to their accretion histories \citep{Roth2016}. For our purposes, we uses the GM technique to decrease the mass of the satellite which exists at $z$ = 0 in P0 and shrank its mass \textit{prior} to when it enters the galaxy at $z$ = 1. To create the modified set of initial conditions, we determined which elements in the linear overdensity field of the initial condition grid map to the particles in the satellite. We then decreased the mean overdensity of these elements in the initial linear vector, all the while, maintaining the mean overdensity of the elements mapping to the main halo to preserve the final mass. We note that the effect of changing the satellite mass for the GM galaxies with BH physics does produces a shift in the dark matter mass of the halo (decreases $\leq$ 20\%). The physical reason for this drop is that AGN feedback suppresses DM accretion following a particularly strong expulsion of gas. Additionally, the inclusion of baryonic physics can result in a decrease in total halo mass \citep{Munshi2013} which helps account for this drop. However, as the difference between our halo masses ($\sim$ 0.2 dex) is much smaller than our stellar masses ($\sim$ 0.2 dex), to a good approximation, we clearly explore the effect of changing M$_{*}$ with fixed M$_{halo}$.

\subsubsection{Galaxies with BH Physics} 

%\vspace{-5 mm}
\begin{table}[ht!] 
\caption{Zoom-In Galaxies Modification} % title of Table
\centering % used for centering table
\begin{tabular}{c c} % centered columns (4 columns)
\hline\hline %inserts double horizontal lines
Sim & Satellite Dark Matter Mass  \\
 & (M$_{\odot}$) at \textit{z} = 1\\ [0.5ex] % inserts table
%heading
\hline % inserts single horizontal line
P0 & 7.3 $\times$ 10$^{10}$ \\ % inserting body of the table
GM1 & 5.9 $\times$ 10$^{10}$ \\
GM2 & 4.0 $\times$ 10$^{10}$ \\ % This used to be GM7
GM3 & 2.5 $\times$ 10$^{10}$ \\ [1ex] % [1ex] adds vertical space; GM3 used to be GM4
\hline %inserts single line
\vspace{0mm}
\end{tabular}
\label{table:satdata} % is used to refer this table in the text
\end{table}

\begin{table*}[ht!] % `*' makes it go across both columns instead of one
\caption{Properties of Zoom-In Galaxies \textit{with BHs} at z = 0.17} % title of Table
\centering % used for centering table
\begin{tabular}{c c c c c c c} % centered columns (4 columns)
\hline\hline %inserts double horizontal lines
Sim & Total Halo Mass  & Total Gas Mass & Total Stellar Mass & CGM Gas Mass & R$_{vir}$ & T$_{vir}$ \\
 & (M$_{\odot}$) & (M$_{\odot}$) & (M$_{\odot}$) & (M$_{\odot}$) & (kpc) & (K) \\ [0.5ex] % inserts table
%heading
\hline % inserts single horizontal line 
P0 & 9.9 $\times$ 10$^{11}$ & 1.1 $\times$ 10$^{11}$ & 5.0 $\times$ 10$^{10}$ & 9.3 $\times$ 10$^{10}$ & 277.0 & 5.5 $\times$ 10$^5$ \\ 
GM1 & 9.7 $\times$ 10$^{11}$ & 9.9 $\times$ 10$^{10}$ & 4.7 $\times$ 10$^{10}$ & 8.5 $\times$ 10$^{10}$ & 274.9 & 5.4 $\times$ 10$^5$ \\
GM2 & 8.1 $\times$ 10$^{11}$ & 6.9 $\times$ 10$^{10}$ & 1.4 $\times$ 10$^{10}$ & 6.9 $\times$ 10$^{10}$ & 259.2 & 4.8 $\times$ 10$^5$ \\ % This used to be GM7
GM3 & 6.6 $\times$ 10$^{11}$ & 5.1 $\times$ 10$^{10}$ & 1.1 $\times$ 10$^{10}$ & 5.1 $\times$ 10$^{10}$ & 241.7 & 4.2 $\times$ 10$^5$ \\ [1ex] % [1ex] adds vertical space; GM3 used to be GM4
\hline %inserts single line
\end{tabular}
\label{table:BHdata} % is used to refer this table in the text
\vspace{-2mm}
\end{table*}

\begin{table*}[ht] % `*' makes it go across both columns instead of one
\caption{Properties of Zoom-In Galaxies \textit{without BHs} at z = 0.17} % title of Table
\centering % used for centering table
\begin{tabular}{c c c c c c c} % centered columns (4 columns)
\hline\hline %inserts double horizontal lines
Sim & Total Halo Mass  & Total Gas Mass & Total Stellar Mass & CGM Gas Mass & R$_{vir}$ & T$_{vir}$\\
 & (M$_{\odot}$) & (M$_{\odot}$) & (M$_{\odot}$) & (M$_{\odot}$) & (kpc) & (K) \\ [0.5ex] % inserts table
%heading
\hline % inserts single horizontal line
P0noBH & 9.8 $\times$ 10$^{11}$ & 8.2 $\times$ 10$^{10}$ & 7.9 $\times$ 10$^{10}$ & 7.5 $\times$ 10$^{10}$ & 276.1 & 5.4 $\times$ 10$^5$ \\ % inserting body of the table
GM1noBH & 9.9 $\times$ 10$^{11}$ & 8.7 $\times$ 10$^{10}$ & 7.4 $\times$ 10$^{10}$ & 8.0 $\times$ 10$^{10}$ & 276.2 & 5.5 $\times$ 10$^5$ \\
GM2noBH & 9.6 $\times$ 10$^{11}$ & 8.8 $\times$ 10$^{10}$ & 7.0 $\times$ 10$^{10}$ & 8.0 $\times$ 10$^{10}$ & 274.0 & 5.3 $\times$ 10$^5$ \\ % This used to be GM7
GM3noBH & 8.4 $\times$ 10$^{11}$ & 7.1 $\times$ 10$^{10}$ & 7.3 $\times$ 10$^{10}$ & 6.4 $\times$ 10$^{10}$ & 261.9 & 4.9 $\times$ 10$^5$ \\ [1ex] % [1ex] adds vertical space; GM3 used to be GM4
\hline %inserts single line
\vspace{5.111mm}
\end{tabular}
\label{table:noBHdata} % is used to refer this table in the text
\end{table*}

At $z$ = 0, our P0 galaxy is a star forming galaxy with a disk (Figure \ref{fig-P0}). P0 has an incoming satellite at $z$ = 0 with an original mass of 7.34 $\times$ 10$^{10}$ M$_{\odot}$ (mass ratio, q = 0.12) \textit{prior} to entering the main halo's virial radius at $z$ $\sim$ 1. For each GM galaxy simulation, we systematically shrink this satellite halo's mass prior to its entry into the main halo (Table \ref{table:satdata}). GM1 results in a similar disked, star forming galaxy, while GM2 and GM3 become quenched at $z$ $\sim$ 1 (Table \ref{table:BHdata}). 

Patient 0 and its 3 GM simulations have mass resolutions of 1.4 $\times$ 10$^{5}$ M$_{\odot}$ and 2.1 $\times$ 10$^{5}$ M$_{\odot}$ for DM and gas particles, respectively. The DM field in these galaxies is simulated at twice the gas resolution to reduce noise in the potential near the galactic center \citep{Pontzen2016} and more accurately trace BH dynamics \citep{Tremmel2015}.

While these GM galaxies are generated using the same method as \cite{Pontzen2016}, their study examines a different set of galaxies. The three galaxies in \cite{Pontzen2016} were run to $z$ = 2 and have M$_{Halo}$ $\sim$ 10$^{12}$ M$_{\odot}$. They each have incoming satellites whose masses are both increased and decreased prior to merging with the main galaxy, as in our galaxies. We note that the genetic modifications performed on the galaxies of \citep{Pontzen2016} were different from the ones implemented here. In their case, it was an enhanced merger (increased satellite's mass) that resulted in a quenched galaxy, rather than a shrunken satellite mass as we implement here. However, in our quenched galaxies, we see that the mass is compensated by faster, early accretion to account for maintaining the main halos' final masses.

\subsubsection{Galaxies without BH Physics}
One key benefit of the individual zoom-in galaxies includes the ability to remove or adjust the parameters affecting our galaxies. This capability allows us to test different theoretical models which would be too computationally expensive to do with a large volume like R25. In particular, we may exploit this utility to understand directly the effects of the SMBH. To isolate the effect of the SMBH on the CGM, all four of the zoom-in simulations (P0 and its 3 GMs) were re-simulated at the same resolution and with all the same physics \textit{excluding} BH formation, feedback, and dynamical friction (Table \ref{table:noBHdata}). Black hole seed formation was disabled and the BH feedback and accretion efficiency parameters set to 0.

\begin{figure}[ht!]
\centerline{\resizebox{1.02\hsize}{!}{\includegraphics[angle=0]{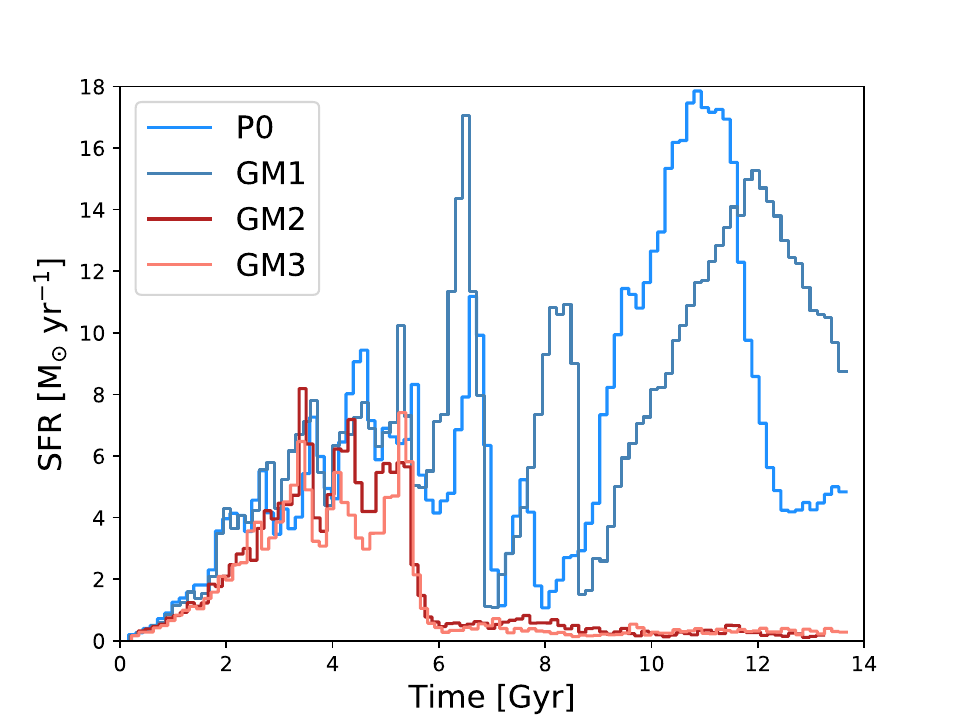}}}
\vspace{1mm}
\centerline{\resizebox{1.02\hsize}{!}{\includegraphics[angle=0]{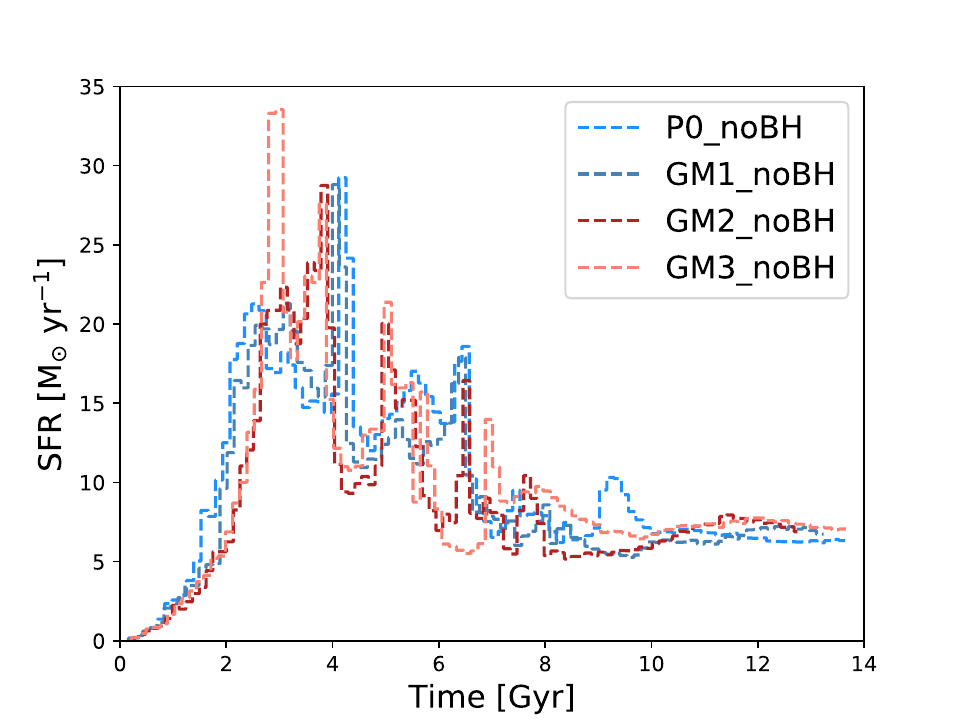}}}
\caption[]{The star formation histories for the zoom-in galaxies: Patient 0 and its 3 GM galaxies with BH physics (\textit{Upper}) and without BH physics (\textit{Lower}). In the galaxies including BH physics, P0 and GM1 remain star forming throughout their histories while GM2 and GM3 become quenched at $z$ $\sim$ 1. Without BH physics, all four galaxies remain star forming until $z$ = 0.}
\label{fig-sfh}
\end{figure}

\begin{figure}[ht!] 
\centerline{\resizebox{1.05\hsize}{!}{\includegraphics[angle=0]{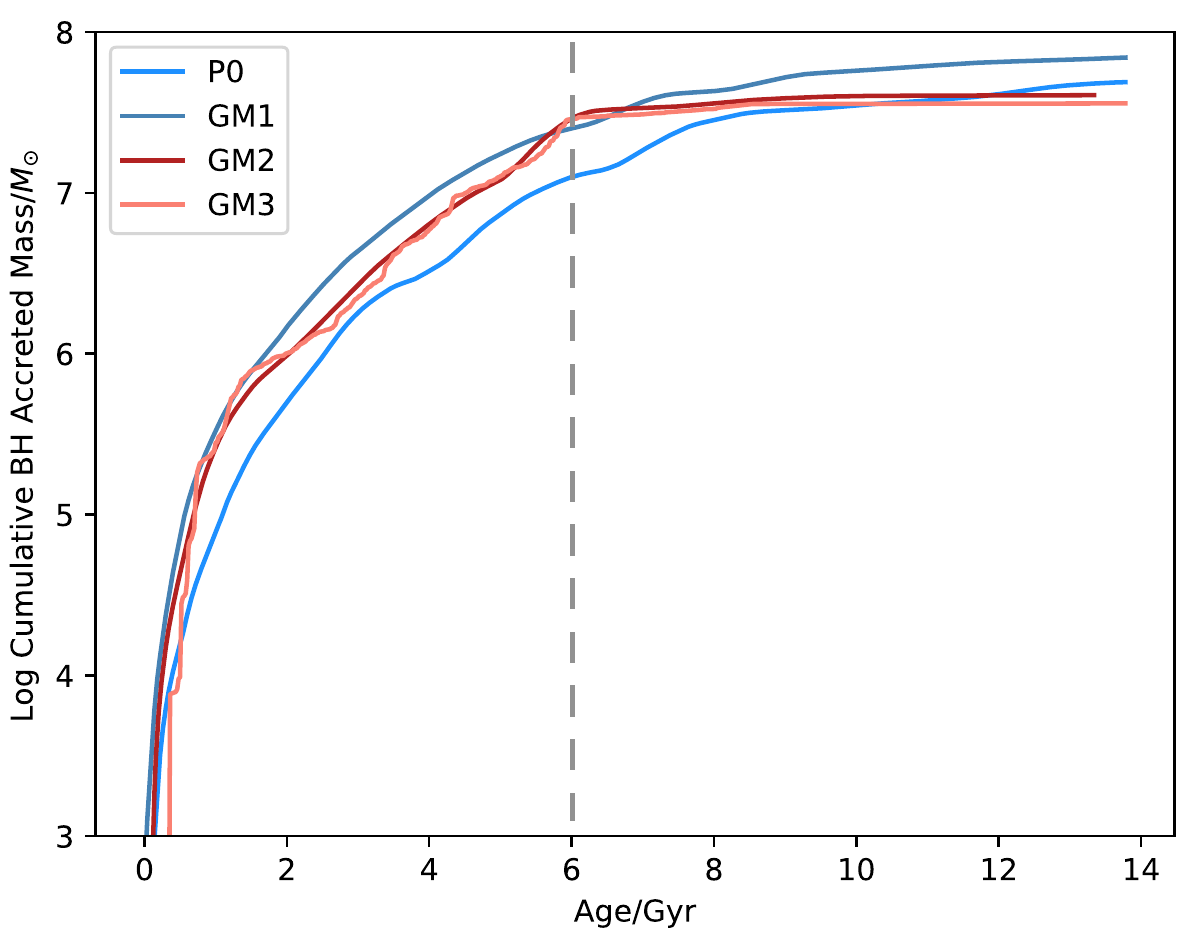}}}
\vspace{2mm}
\centerline{\resizebox{1.07\hsize}{!}{\includegraphics[angle=0]{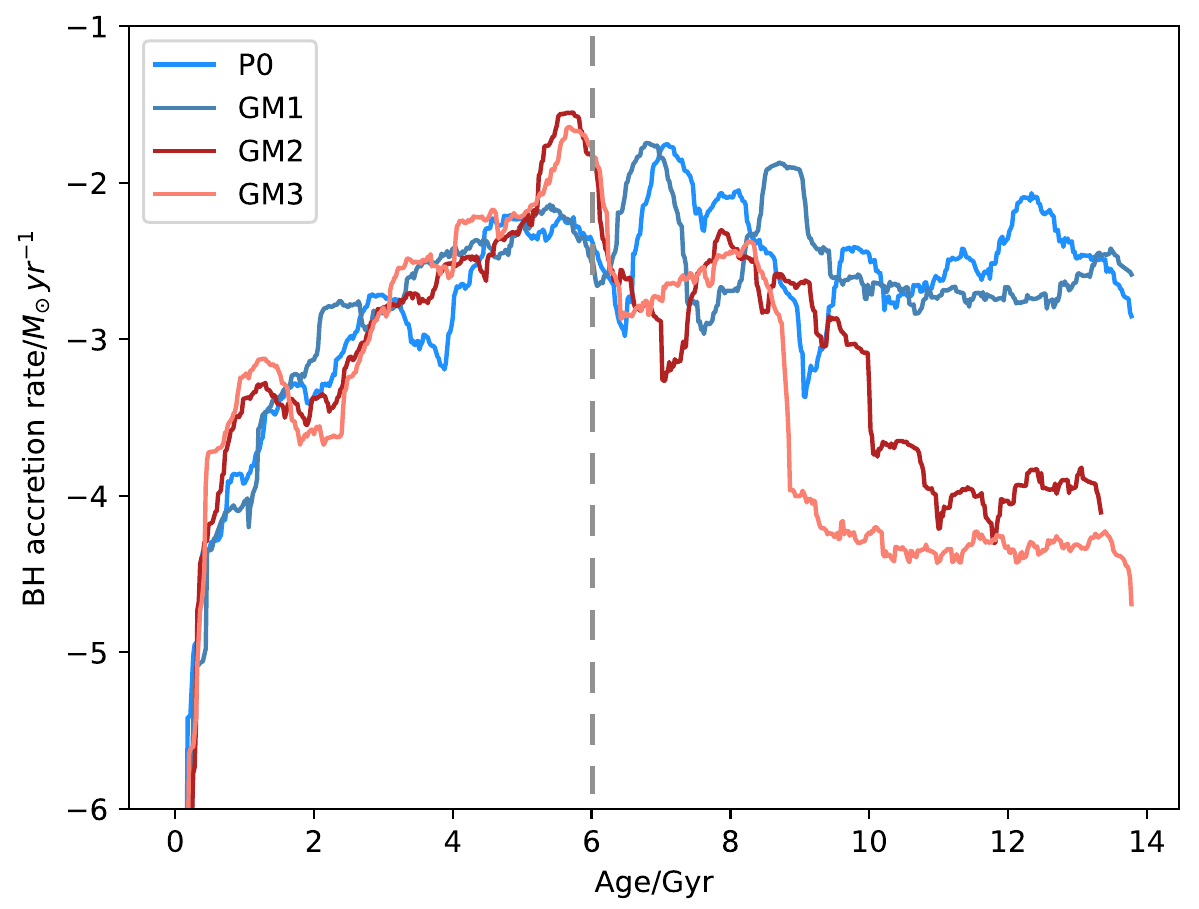}}} 
\caption[]{SMBH accreted mass (\textit{Upper}) and SMBH accretion rates (\textit{Lower}) for our 4 zoom-in galaxies. Colors as in Figure \ref{fig-sfh}. The accreted mass of all the SMBHs are comparable. However, both quenched galaxies also have a sharp peak in accretion rate around the time of the most significant merger (z $\sim$ 1, t $\sim$ 6 Gyr), indicated by the dashed grey line.}
\label{fig-bh}
\end{figure}

\begin{figure}[ht]
\centerline{\resizebox{01.06\hsize}{!}{\includegraphics[angle=0]{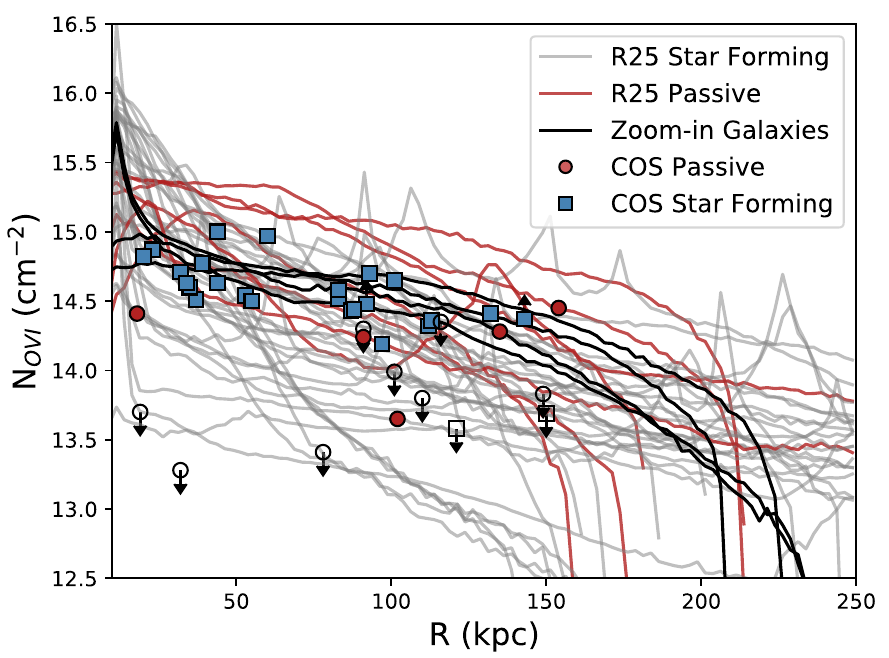}}}
\caption[]{Mean column densities of \ion{O}{6} as a function of radius for all 39 of the galaxies in R25 which fall within the COS-Halos stellar mass range and our family of zoom-in galaxies. All galaxies are examined at $z$ = 0.17. Solid grey and red lines indicate R25 star forming and quenched galaxy column densities, respectively. Solid black lines describe the column densities of our four zoom-in galaxies. Filled circles and squares indicate star forming and passive galaxies from the COS-Halos Survey dataset. Unfilled markers indicate upper limits.}
\label{fig-ROM_GMs_NOvi}
\end{figure}

\subsubsection{Quenching in GM2 and GM3} \label{subsec-quench}
The top panel of Figure \ref{fig-sfh} shows star formation histories of the four zoom-in galaxies with BH physics included. P0 and GM1 are in light and dark blue, respectively, while GM2 and GM3 are shown similarly in dark and light red. Their star formation histories demonstrate that, unlike P0 and GM1 which remain star forming throughout their history, GM2 and GM3 become quenched at $z$ $\sim$ 1. Contrastingly, the lower panel of Figure \ref{fig-sfh} shows the star formation histories of the four zoom-in galaxies without BH physics and all four of their histories remain star forming and are fairly similar. The immediate quenching seen in the upper panel for GM2 and GM3, which occurs just after the merger of the satellite with the main halo, \textit{does not} take place in the simulations of GM2 and GM3 without BH physics, consistent with \cite{Pontzen2017}. Significant outflows after the time of the merger (z $\sim$ 1) result in the quenching we see in GM2 and GM3; however, we leave a detailed treatment of the mechanisms driving these outflows to future work.  The stark differences between the GM2 and GM3 galaxies with and without BHs imply that some interplay between the satellite's mass and the SMBH feedback must play a pivotal role in quenching these galaxies so thoroughly. 

\cite{Pontzen2017} previously explored the relationship between BH feedback and mergers and its effect on quenching, using the same genetic modification technique as we use for the GM galaxies in our study. They determine that SMBH feedback is critical to quenching a galaxy, which is consistent with our finding that quenched galaxies arise only in simulations that include SMBHs (Figure \ref{fig-sfh}). \cite{Pontzen2017} argue that mergers can disrupt the cold disk of the galaxy, allowing SMBH feedback to more strongly suppress star formation in the disk and keep the galaxy in a state of quiescence. Mergers have also been shown to help funnel gas into the region of the SMBH allowing for more direct accretion \citep{Richards2006,Hopkins2010,Nelson2013,Sanchez2017}.

We further examine the effects of the BH by looking to the accreted mass and accretion rates of the BHs. The upper panel in Figure \ref{fig-bh} (colors as in Figure \ref{fig-sfh}) shows the cumulative accreted SMBH mass as a function of time. Here we see that the accreted mass growth in the quenched galaxies, GM2 and GM3, is similar to that of the star forming galaxies. However, more significant differences arise in the lower panel of Figure \ref{fig-bh}, which depicts the SMBH accretion rates as a function of time. From this figure, we can see an increase of accretion occurs for both quenched galaxies near the time of the merger (z $\sim$ 1, t $\sim$ 6 Gyr). In particular, for the two quenched galaxies GM2 and GM3 (shown in dark and light red, respectively), we see that the accretion rate peaks about a Gyr earlier than for the star forming galaxies. The accretion rate in the quenched galaxies continues to drop after this point, while the SMBH in each star forming galaxy continues to accrete. Although the BH's activity and growth are not directly affected by the changing mass of the incoming satellite, together the modified satellite mass and effect of the BH make a significant impact on the star formation history of the galaxy. Thus, while the peak accretion rates are similar in quenched and unquenched cases, the resulting energy couples differently to the galaxies and only in the latter case do they lead to a reduction in later inflows.

This set of galaxies was produced from very similar initial conditions and therefore have near-identical large scale filamentary feeding. However, they illustrate very different star formation and accretion histories and allow us to directly examine how assembly history may imprint itself on the CGM. Additionally, they allow us to concretely confirm the result of \cite{Pontzen2017} that the effect of a SMBH, while not the only requisite, is \textit{vital} to the quenching process in galaxies. 

% ### END OF SIMULATION PARAMETERS ####
% #####################################

\begin{figure}[ht]
\centerline{\resizebox{1.\hsize}{!}{\includegraphics[angle=0]{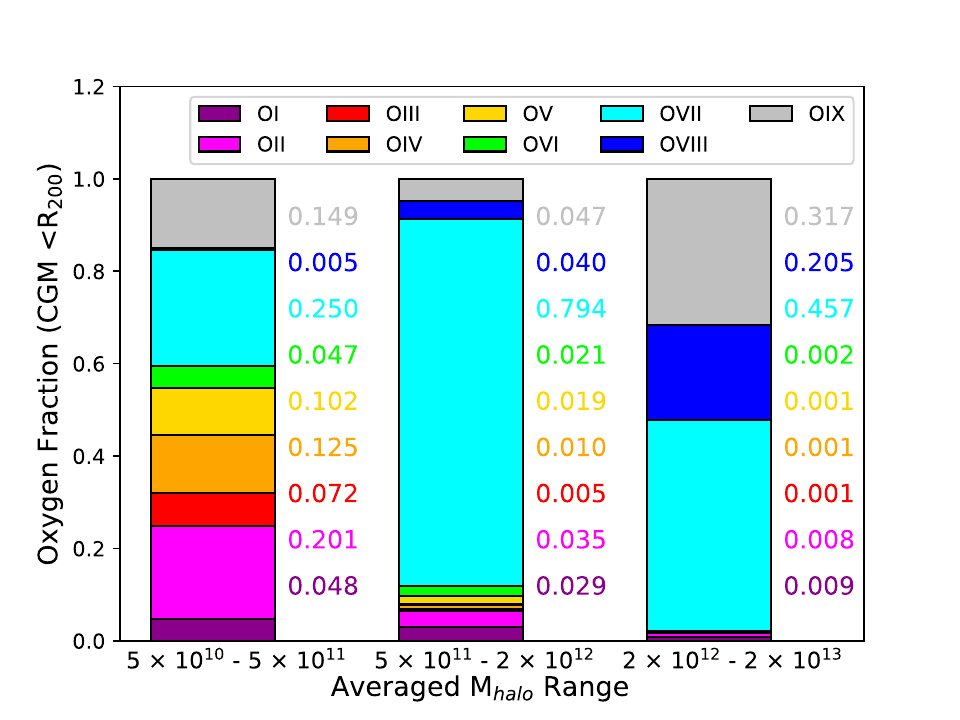}}}
\caption[]{Average oxygen ion fractions in the CGM of R25 within 3 M$_{halo}$ range bins: 5 $\times$ 10$^{10}$ \textemdash 5 $\times$ 10$^{11}$, 5 $\times$ 10$^{11}$ \textemdash 2 $\times$ 10$^{12}$, and 2 $\times$ 10$^{12}$ \textemdash 2 $\times$ 10$^{13}$. \ion{O}{6} is shown by green bars. The individual ion fractions are given in their corresponding colors to the right of each bar, ascending in order from least to most ionized such that \ion{O}{6} is fourth ionization fraction from the top. The average \ion{O}{6} fraction decreases as halo mass increases.}
\label{fig-oppenheimer}
\end{figure}

\begin{figure}[ht]
\centerline{\resizebox{1.1\hsize}{!}{\includegraphics[angle=0]{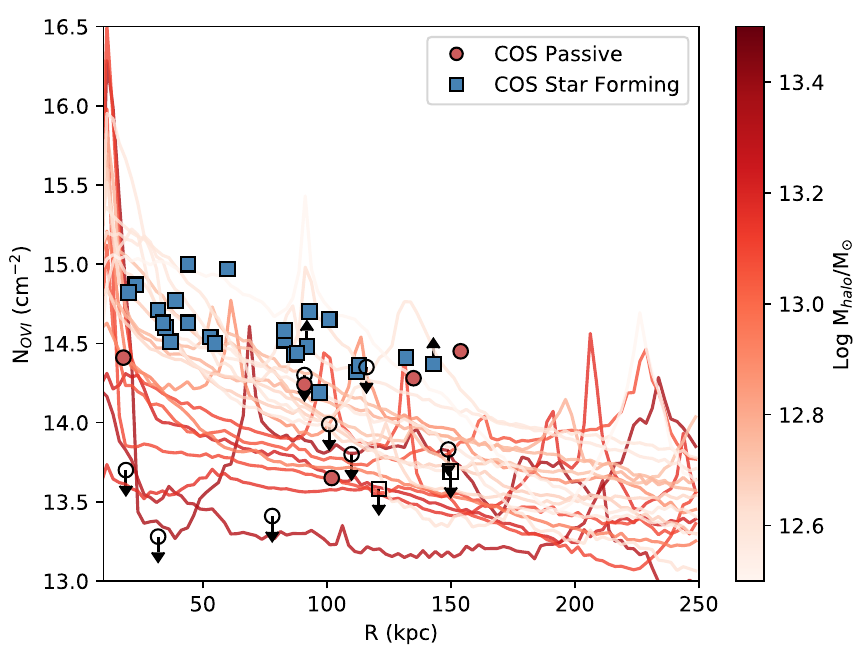}}}
\caption[]{Column density profiles of \ion{O}{6} in the high mass (M$_{vir}$ \textgreater \ 2 $\times$ 10$^{12}$ M$_{\odot}$) galaxies of R25 at z = 0.17.}
\label{fig-highmass_Novi}
\end{figure}

\begin{figure}[ht]
\centerline{\resizebox{1.1\hsize}{!}{\includegraphics[angle=0]{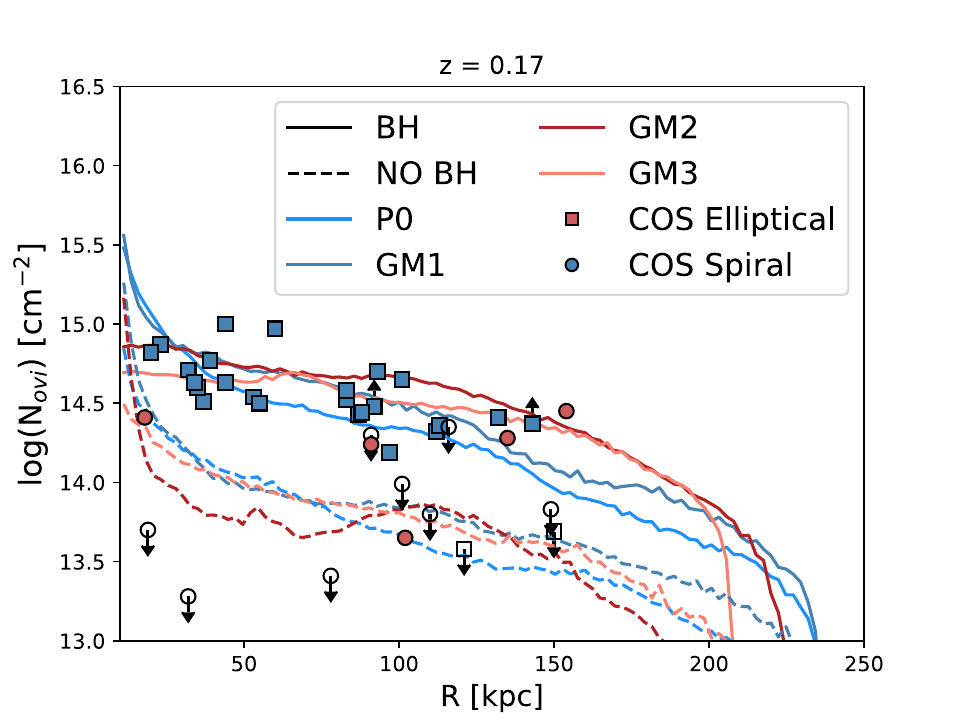}}}
\caption[]{Column density profiles of \ion{O}{6} in our 4 zoom-in galaxies with (solid lines) and without (dashed lines) BH physics. P0 and GM1, our two star forming galaxies are marked in light blue and dark blue, respectively. Our quenched galaxies, GM2 and GM3, are labeled in dark red and light red, respectively. These column densities show that the BH is essential to shaping the \ion{O}{6} in the CGM of star forming and passive galaxies alike.}
\label{fig-GMs_NOvi}
\end{figure}

\begin{figure*}[ht]
\centerline{\resizebox{0.45\hsize}{!}{\includegraphics[angle=0]{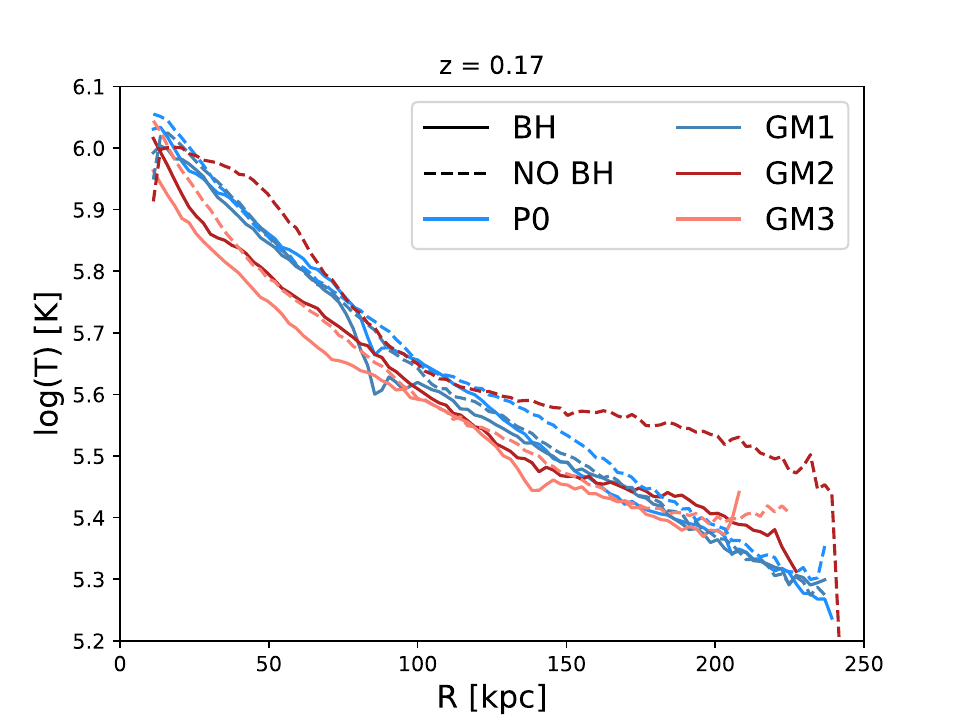}}
\resizebox{0.46\hsize}{!}{\includegraphics[angle=0]{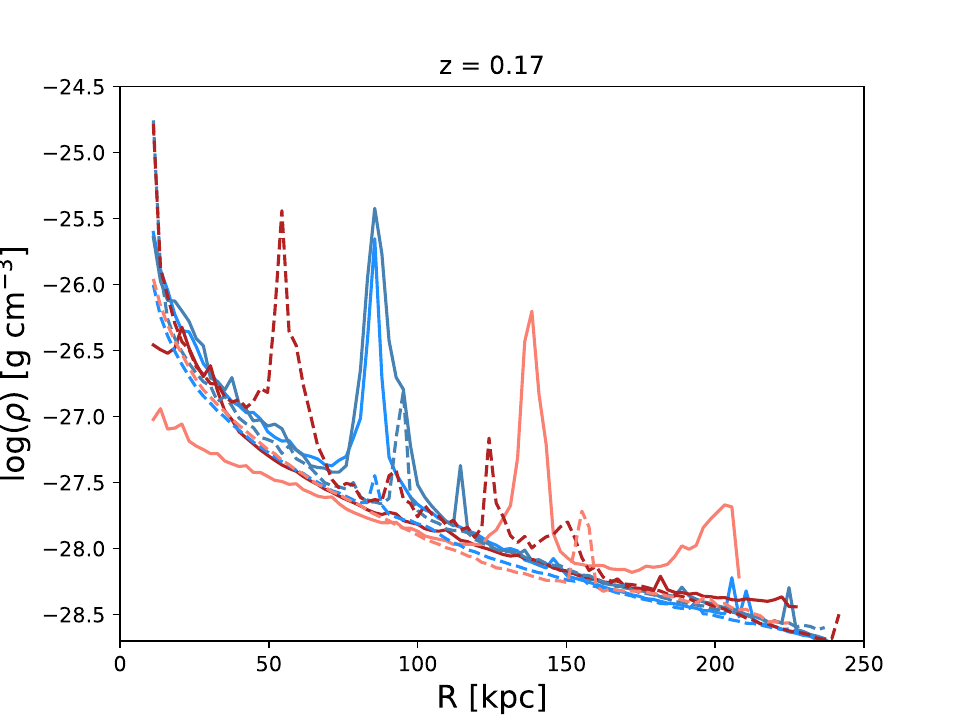}}}
\centerline{\resizebox{0.46\hsize}{!}{\includegraphics[angle=0]{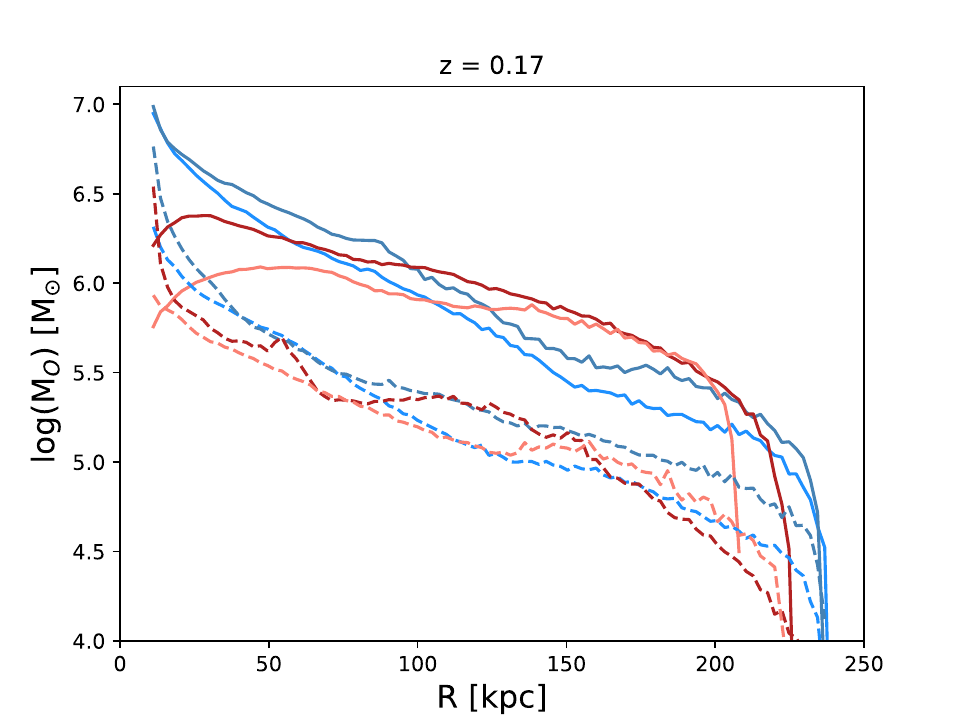}}
\resizebox{0.465\hsize}{!}{\includegraphics[angle=0]{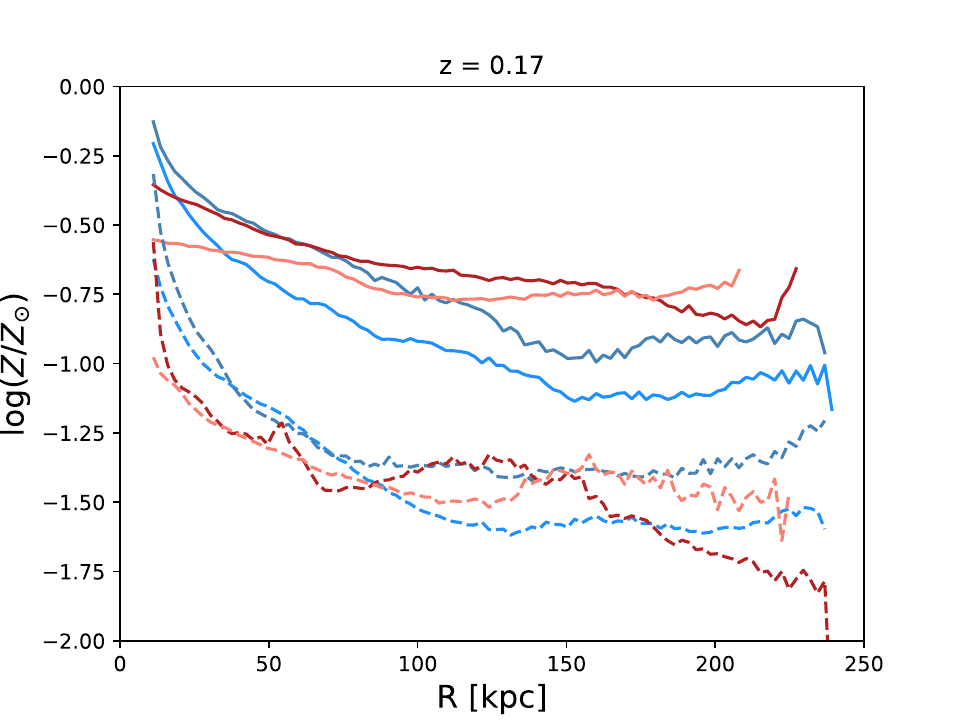}}}
\caption[]{\textit{Clockwise from Upper Left:} Temperature, density, metallicity, and total oxygen mass profiles of the CGM of our 4 zoom-in galaxies with and without BH physics at $z$ = 0.17, the average redshift of COS-Halos. Colors and linestyles as in Figure \ref{fig-GMs_NOvi}. Solid and dashed lines designate simulations with and without BH physics, respectively. }
\label{fig-GMs_profiles}
\end{figure*}

\begin{figure}[ht]
\centerline{\resizebox{1.05\hsize}{!}{\includegraphics[angle=0]{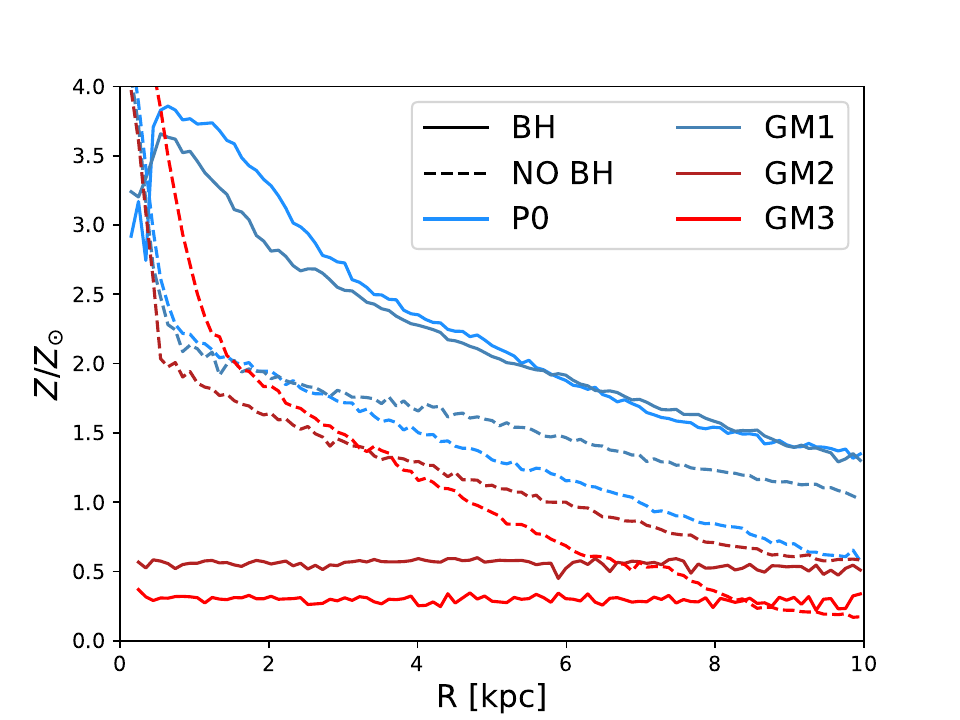}}}
\caption[]{Metallicity profile of the gas within the disk of our 4 zoom-in galaxies with and without BH physics. Colors and line styles as in Figure \ref{fig-GMs_NOvi}. Without the BH physics, metals remain trapped near the center of the disk with no mechanism to propagate out into the CGM.}
\label{fig-disk_Z}
\end{figure}

\begin{figure*}[p!]
\centerline{\resizebox{0.42\hsize}{!}{\includegraphics[angle=0]{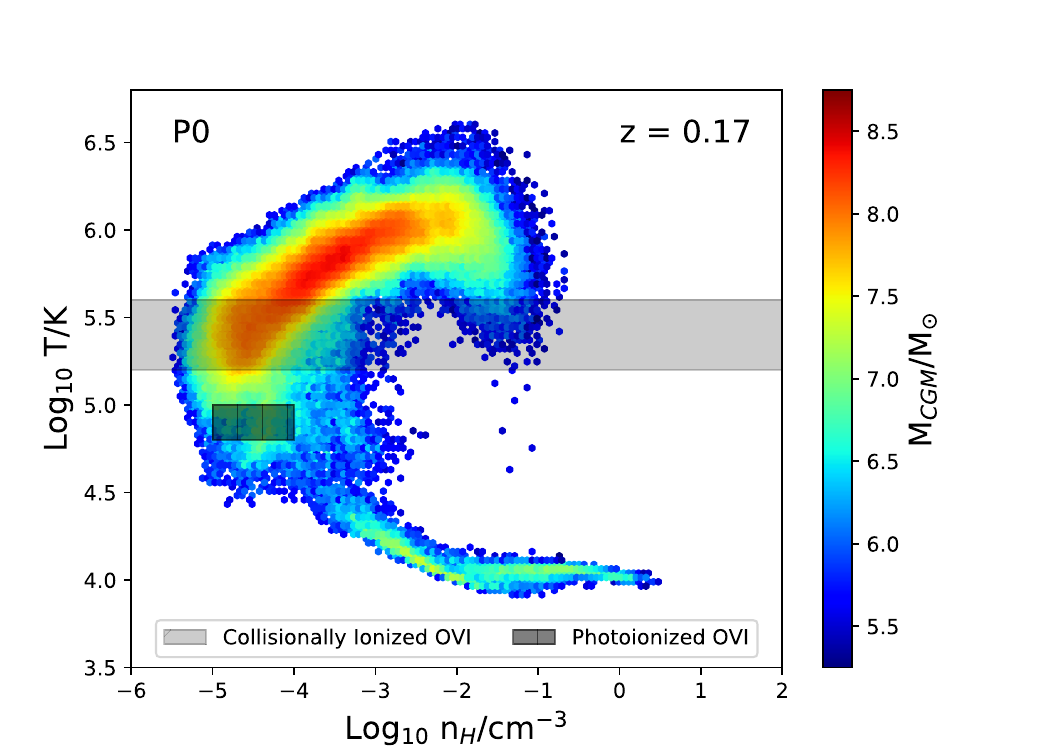}}
\resizebox{0.42\hsize}{!}{\includegraphics[angle=0]{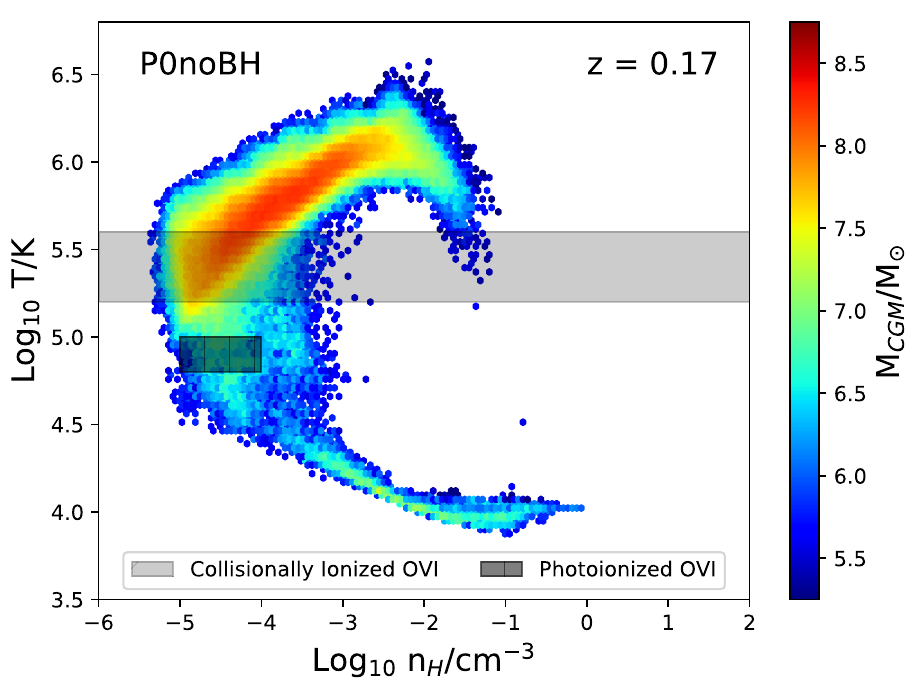}}}
\centerline{\resizebox{0.42\hsize}{!}{\includegraphics[angle=0]{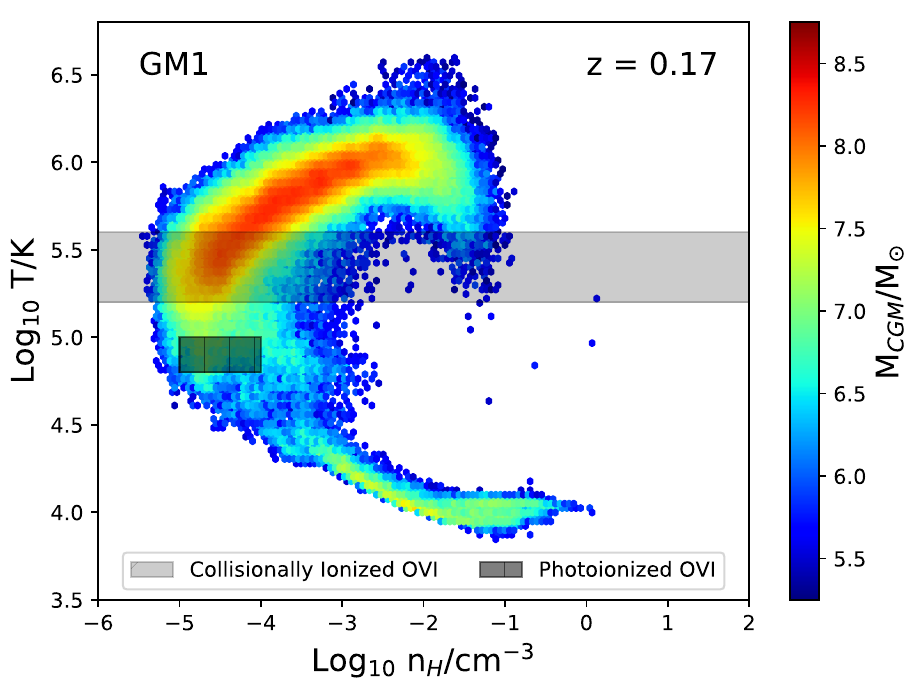}}
\resizebox{0.42\hsize}{!}{\includegraphics[angle=0]{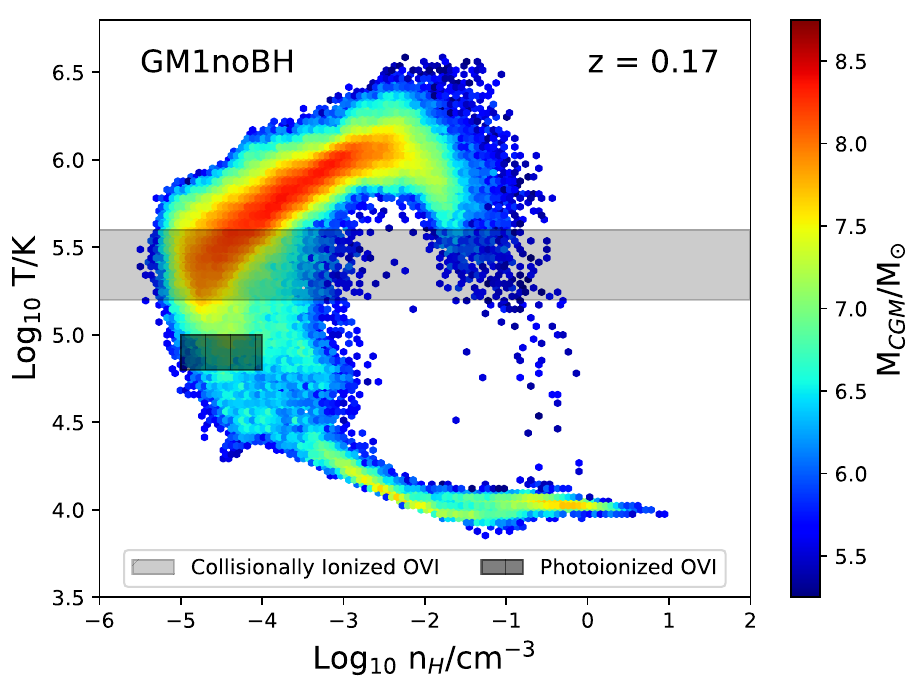}}}
\centerline{\resizebox{0.42\hsize}{!}{\includegraphics[angle=0]{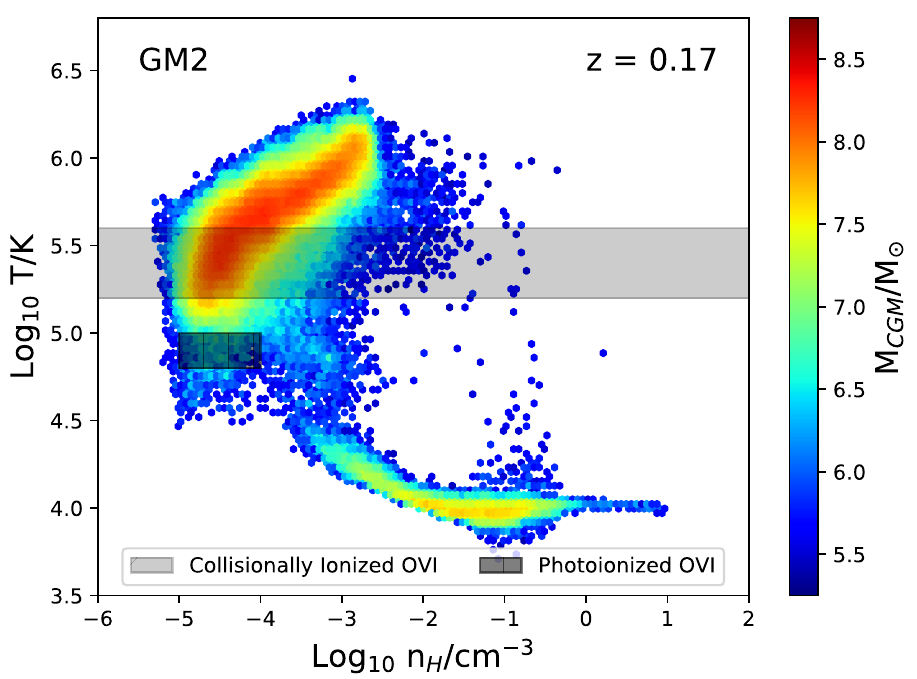}}
\resizebox{0.42\hsize}{!}{\includegraphics[angle=0]{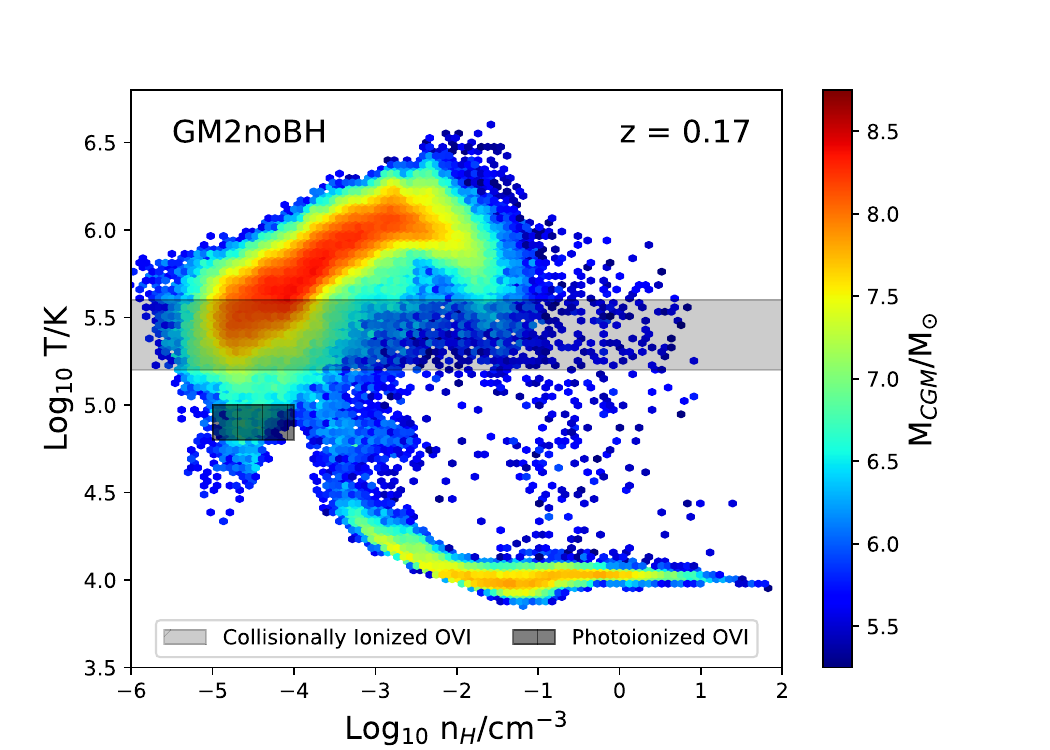}}}
\centerline{\resizebox{0.42\hsize}{!}{\includegraphics[angle=0]{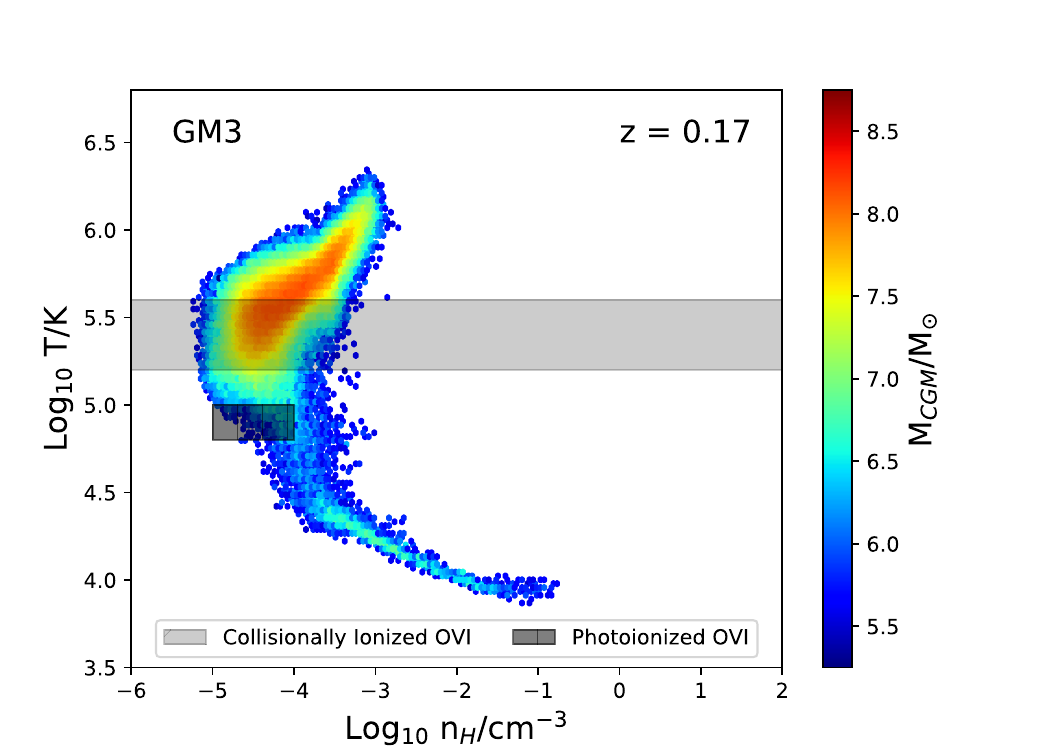}}
\resizebox{0.42\hsize}{!}{\includegraphics[angle=0]{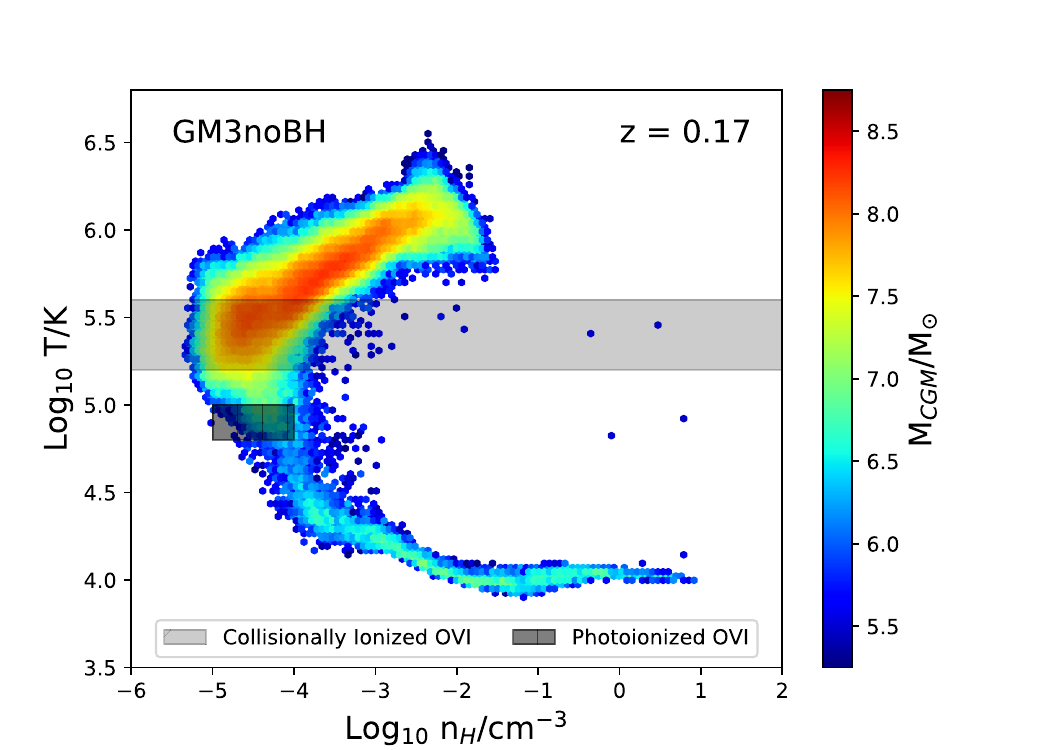}}}
\caption[]{Phase diagrams of the temperature and density of the two star forming zoom-in galaxies, P0 (\textit{Top row}) and GM1 (\textit{Second row}), and the two quenched galaxies, GM2 (\textit{Third row}) and GM3 (\textit{Bottom row}). The phase diagrams of galaxies with BH hole physics vary quite widely between the star forming (P0 and GM1) and quenched cases (GM2 and GM3), particularly in the highest temperature and density gas. However, the phase diagrams of the galaxies without BH physics appear more similar, as are their star formation histories. Semi-transparent light and dark gray boxes span the region of collisionally and photo-ionized \ion{O}{6} as temperature and density regions where fractions of \ion{O}{6} are larger than 0.05 \%.}
\label{phasediagrams}
\end{figure*}

\begin{figure*}[p!]
\centerline{\resizebox{0.385\hsize}{!}{\includegraphics[angle=0]{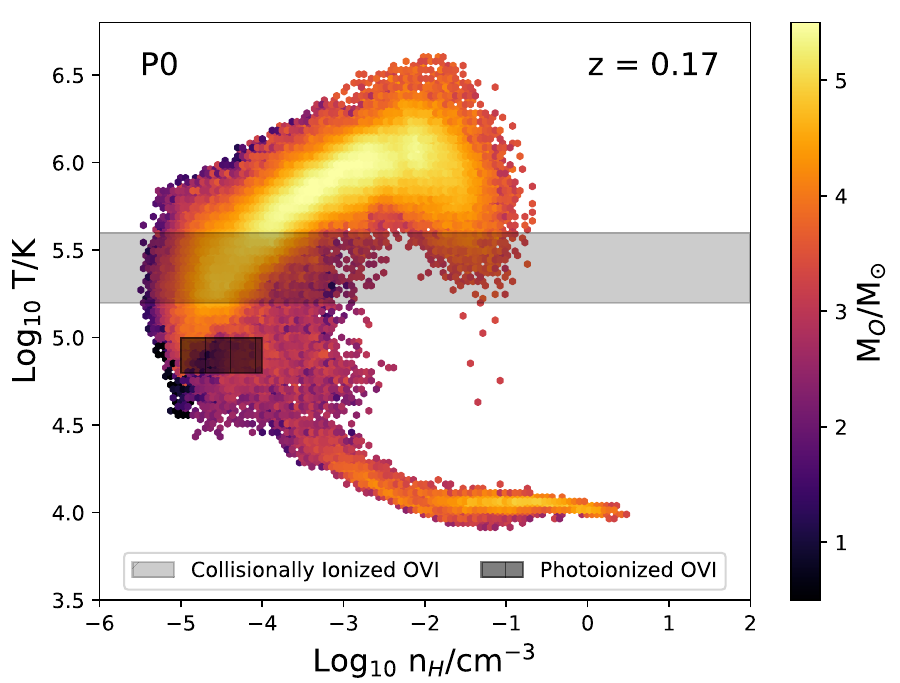}}
\resizebox{0.382\hsize}{!}{\includegraphics[angle=0]{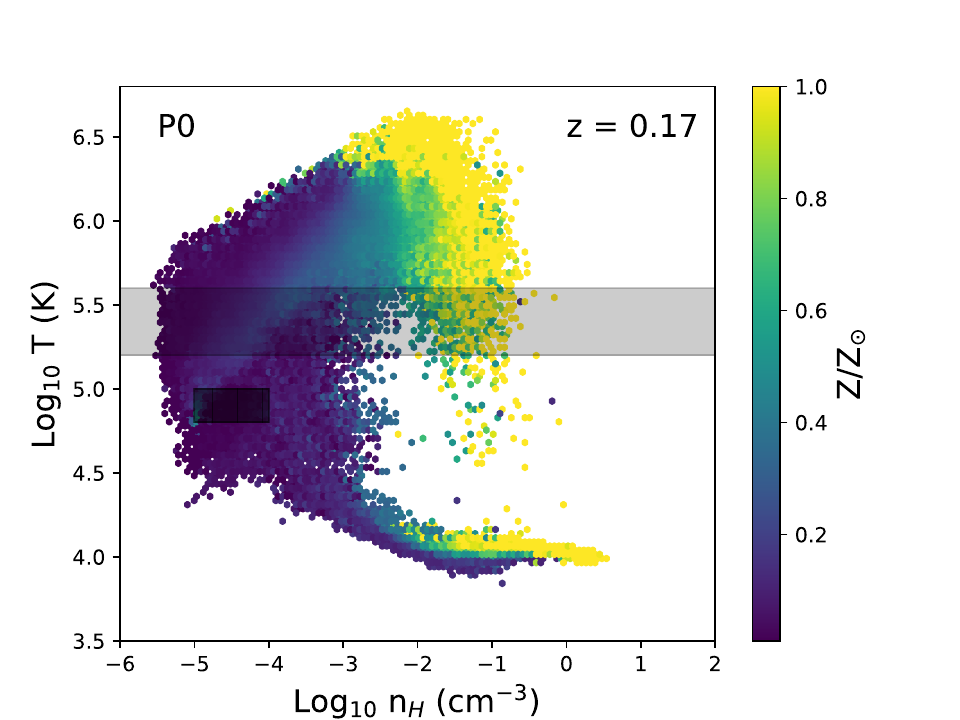}}
\resizebox{0.383\hsize}{!}{\includegraphics[angle=0]{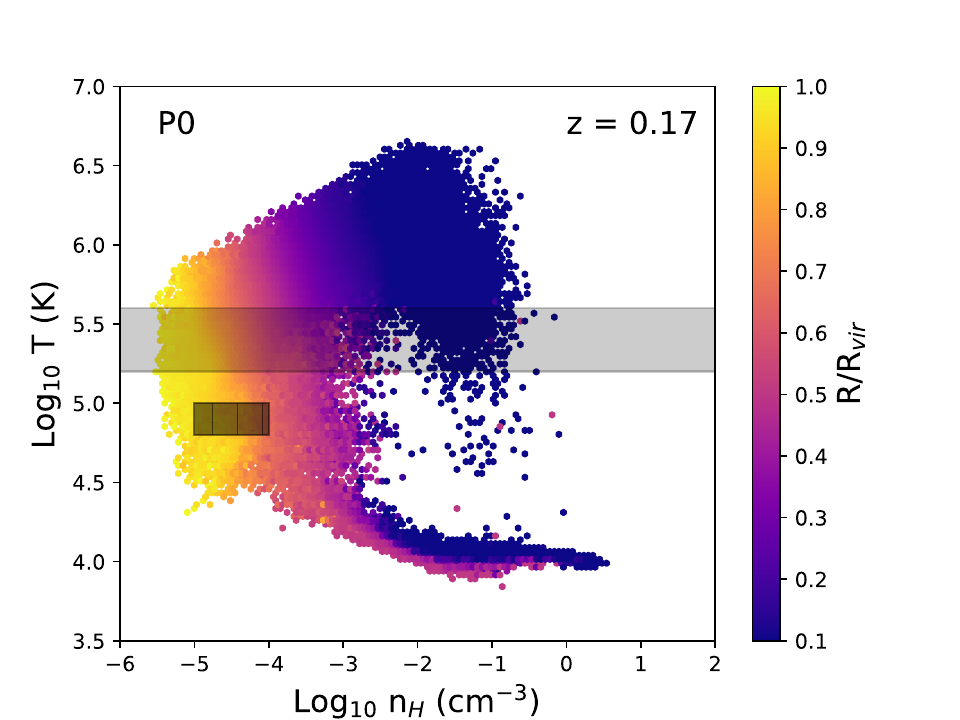}}}
\centerline{\resizebox{0.385\hsize}{!}{\includegraphics[angle=0]{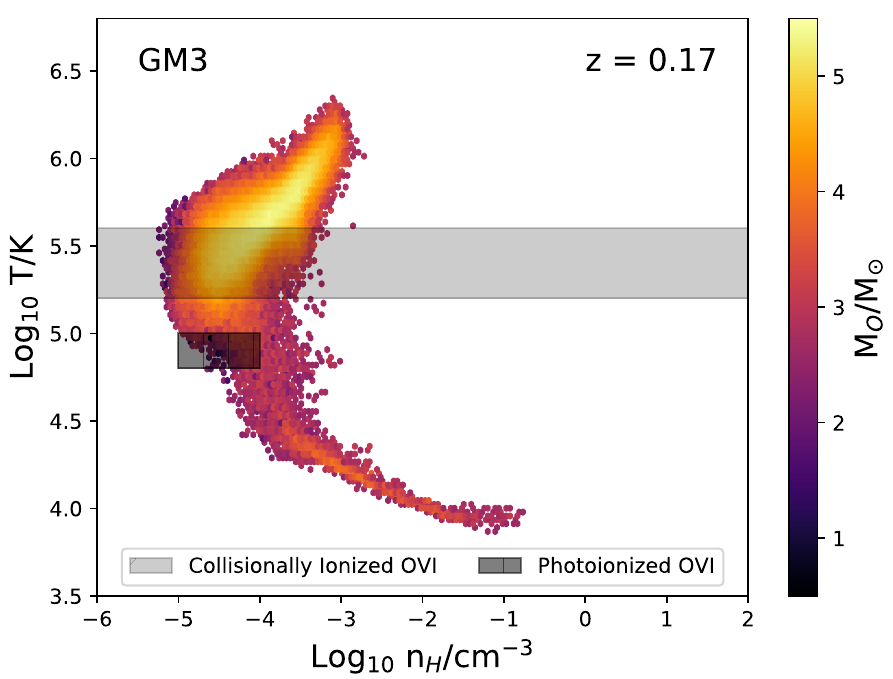}}
\resizebox{0.382\hsize}{!}{\includegraphics[angle=0]{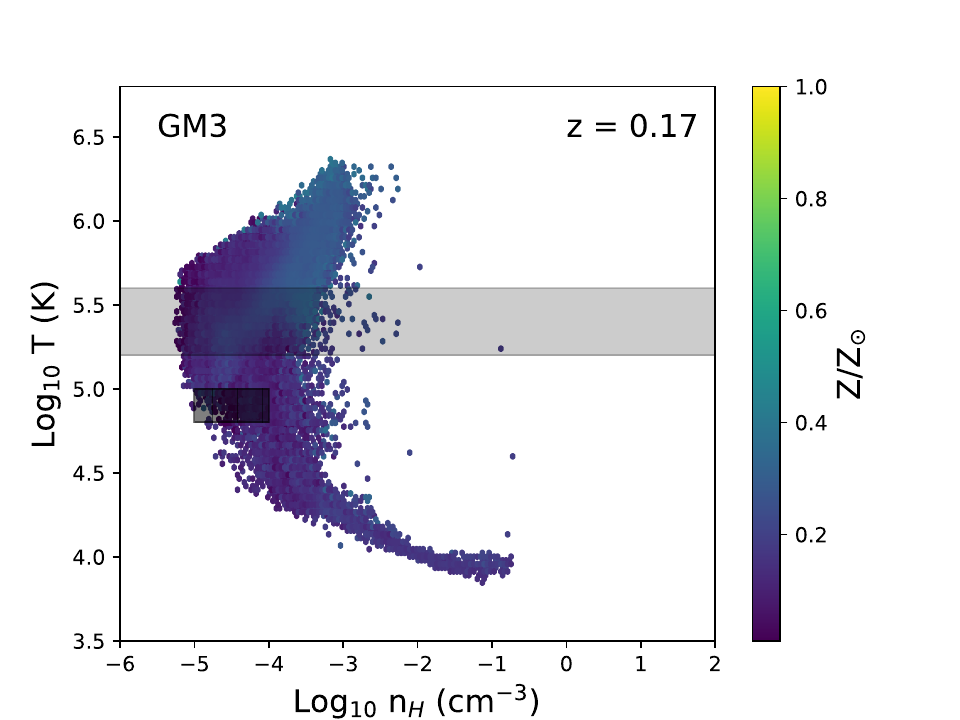}}
\resizebox{0.383\hsize}{!}{\includegraphics[angle=0]{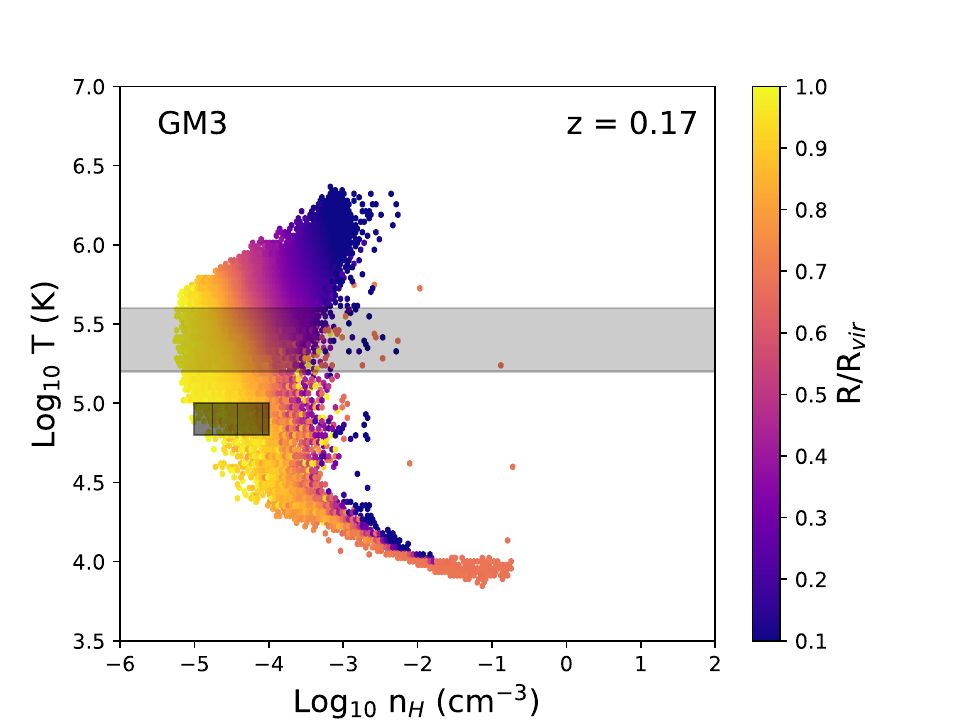}}}
\centerline{\resizebox{0.385\hsize}{!}{\includegraphics[angle=0]{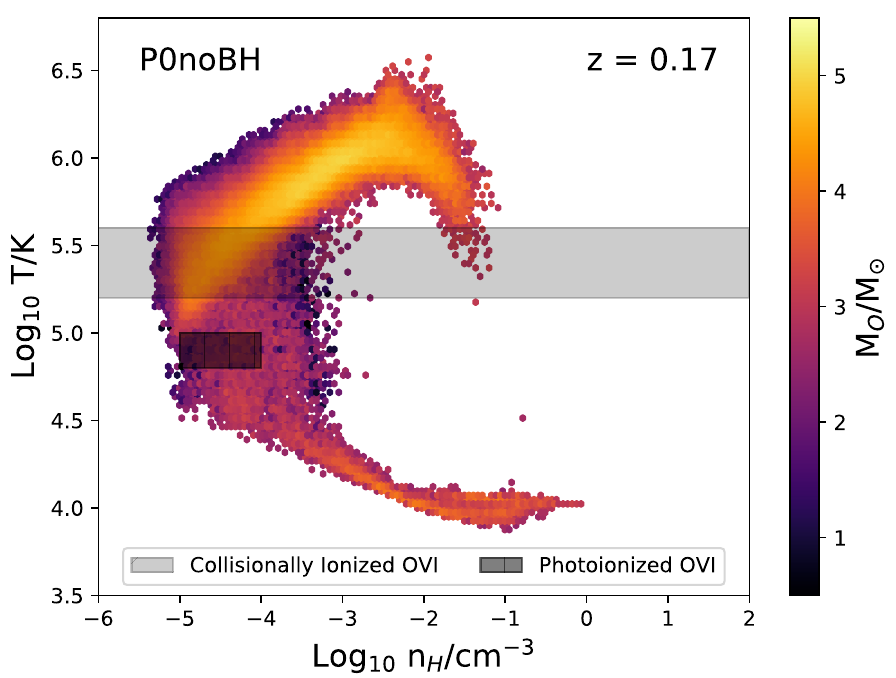}}
\resizebox{0.382\hsize}{!}{\includegraphics[angle=0]{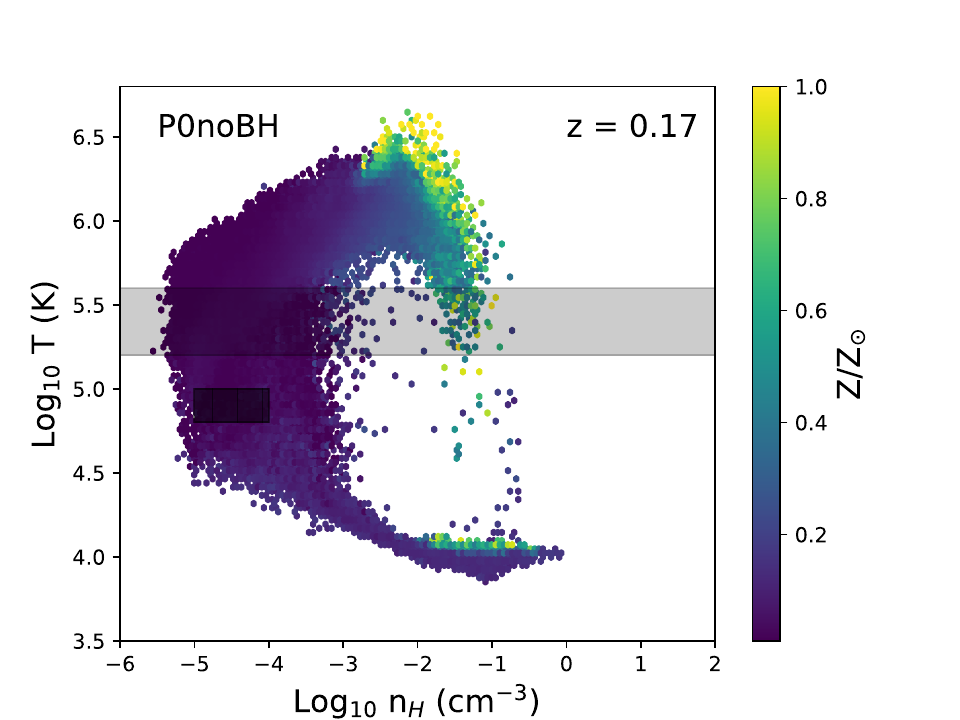}}
\resizebox{0.383\hsize}{!}{\includegraphics[angle=0]{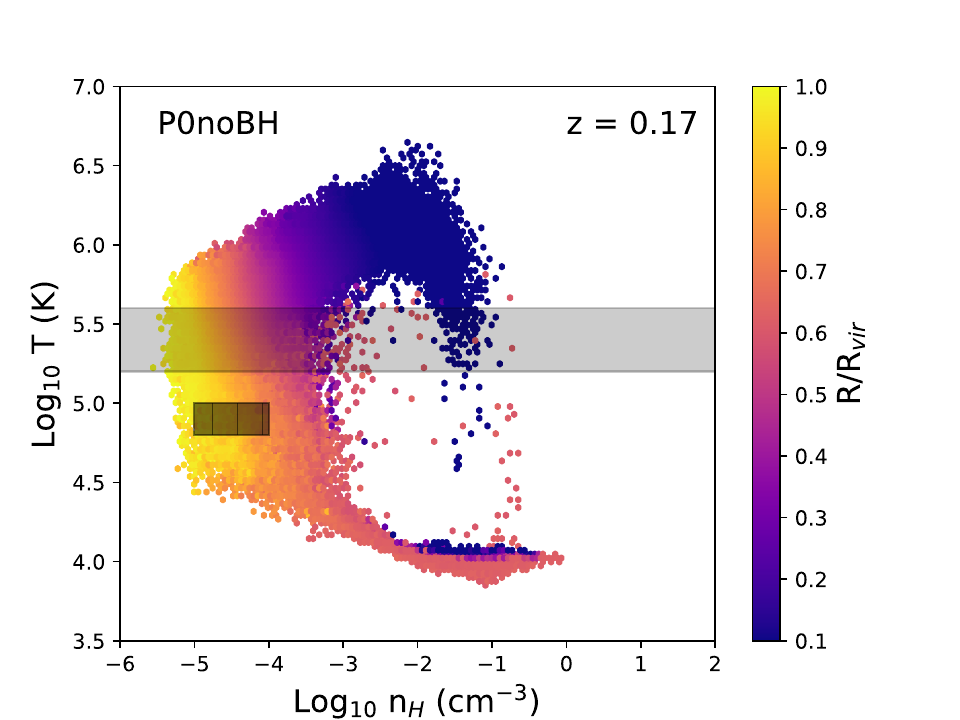}}}
\centerline{\resizebox{0.385\hsize}{!}{\includegraphics[angle=0]{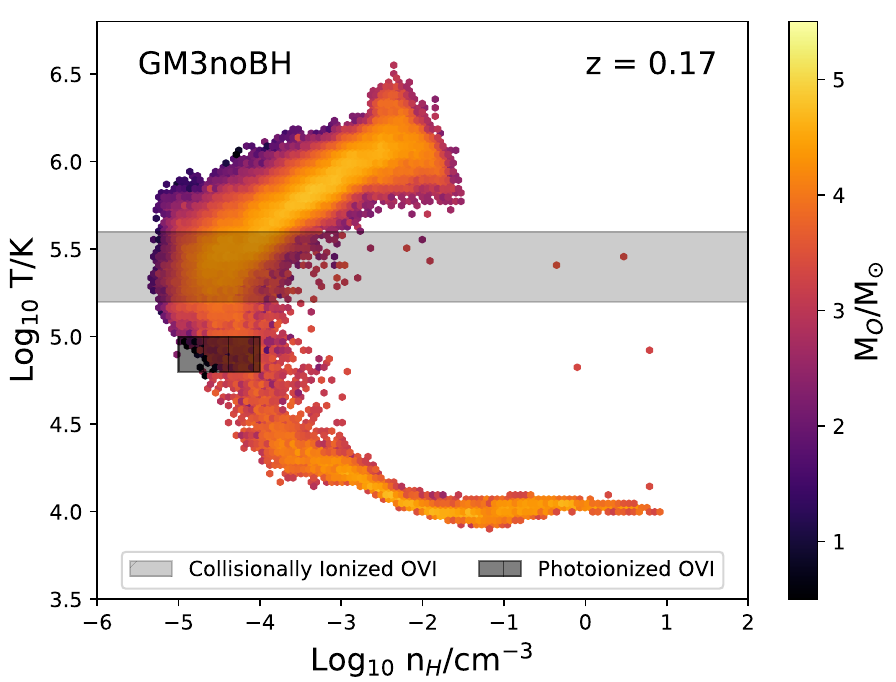}}
\resizebox{0.382\hsize}{!}{\includegraphics[angle=0]{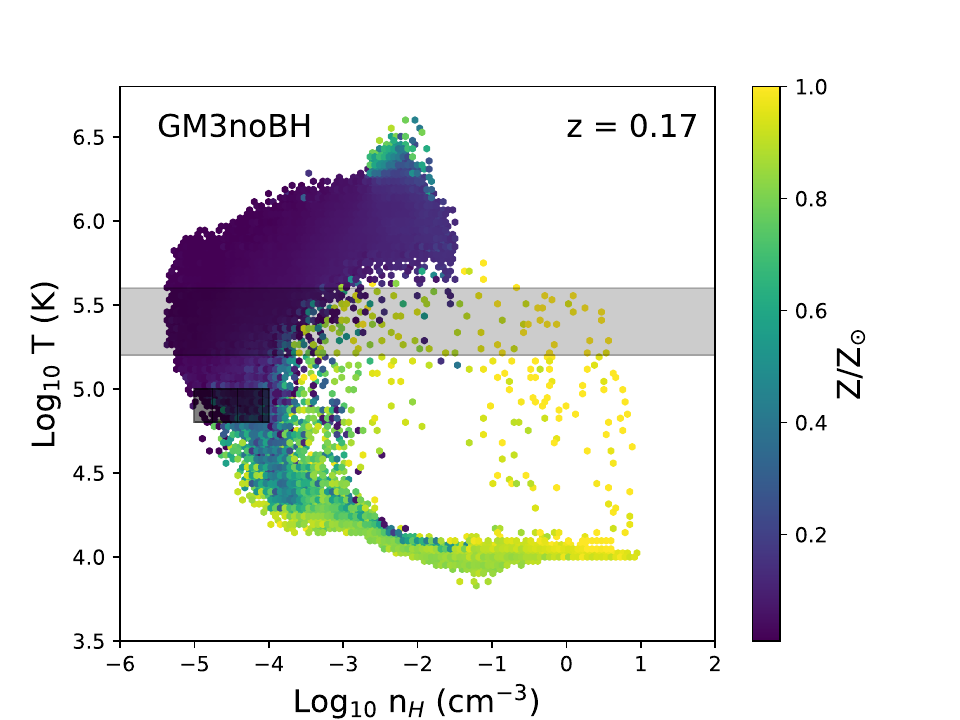}}
\resizebox{0.383\hsize}{!}{\includegraphics[angle=0]{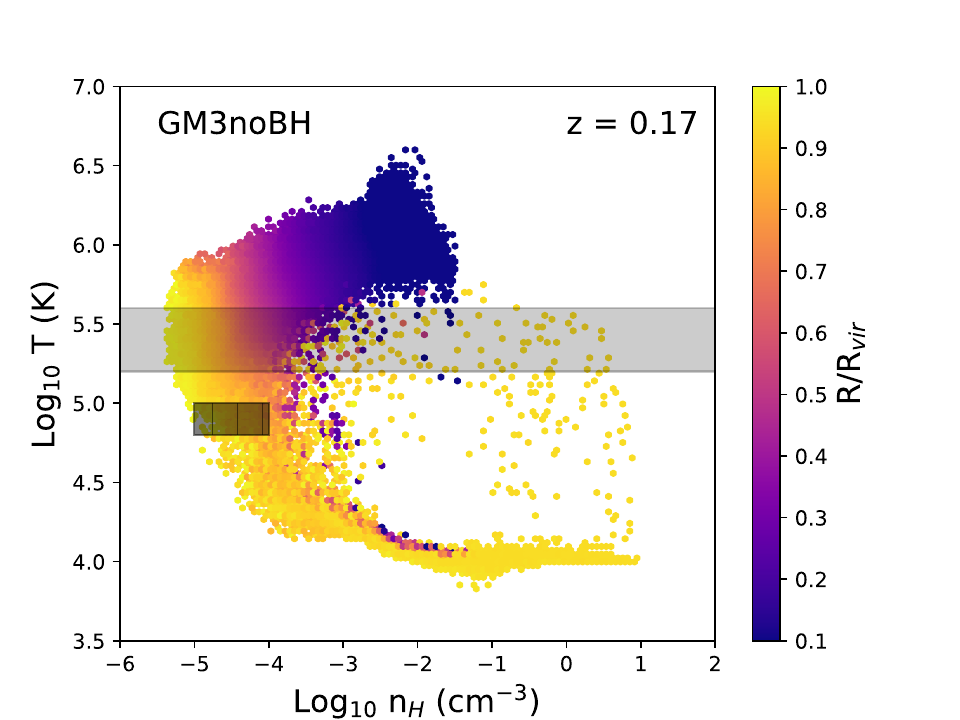}}}

\caption[]{Phase diagrams of the temperature and density of the star forming, P0, and quenched, GM2, with BH physics (\textit{Top two rows} and the same two galaxies without (\textit{Lower two rows}). \textit{Left:} The phase diagrams of these galaxies weighted by the total oxygen mass in each bin. \textit{Middle:} The same phase diagram showing temperature and density, however, the colorbar is weighted by the average metallicity of the gas in each bin. We note that the high density, high temperature gas we see in the star forming P0, is also the highest metallicity gas in the CGM. \textit{Right:} Similarly, a phase diagram with the colorbar now weighted by the average distance from the center of the galaxy of the gas particles in each bin.}
\label{figure:phasediagrams_z_R}
\end{figure*}

% #####################################
% ############# RESULTS ###############
% #####################################
\section{Results}
\label{sec-results}

With the simulations we've described, we examine the effects of stellar evolution and SMBH feedback on setting the contents and physical state of the CGM in MW-mass galaxies. Individual halos in the {\sc Romulus25} cosmological volume and in the individual zoom-in galaxies are extracted using the Amiga Halo Finder (AHF) \citep{Knollmann2009} and central SMBH positions and velocities are defined relative to the center position and inner 1 kpc center-of-mass velocity of their host halo, respectively. All zoom-in galaxies have their most major merger occurring at $z$ $\sim$ 1 (mass ratio = M$_{halo}$/M$_{sat}$, q \textless 10) and an additional merger occurs (q $\sim$ 10) close to $z$ = 0.2, though this time varies slightly across the simulations (See Section \ref{Result:metalsbyBH}).
% their modified satellite halo still present at $z$ = 0. An

The CGM of each individual galaxy halo (within the 39 selected R25 galaxies and our zoom-ins) is defined as the mass enclosed within the virial radius, but further than a spherical radius of 10 kpc away from the center. While the genetic modification process results in galaxies with similar final masses in the absence of strong ejective feedback, we find that the mass of the CGM correlates with the mass of the halo when BH physics is included. P0, which results in the most massive halo at $z$ $\sim$ 0, has the most mass in its CGM, while GM3 results in the least massive CGM mass and halo mass (Table \ref{table:BHdata}).

\subsection{\ion{O}{6} as a Tracer of Virial Temperature Material}
\label{Result:OviasTtracer}
Column densities of \ion{O}{6} are calculated using the analysis software Pynbody \citep{pynbody}. Oxygen enrichment from supernovae and winds is traced throughout the integration of the simulation and ionization states are calculated during post-processing, assuming optically thin conditions, a \cite{Haardt2012} ultraviolet radiation field at $z$ = 0, and collisional ionization equilibrium. Recent papers have raised concerns that this UV background is too weak \citep{Kollmeier2014,Shull2015}. However, as we will soon demonstrate, the \ion{O}{6} in our simulations is predominantly collisionally ionized, so our choice of UV background does not affect our results. We use the CLOUDY software package \citep{Stinson2012,Ferland2013} to create models with varying temperature, density, and redshift to determine \ion{O}{6} fractions for all the gas in each simulated galaxy. Figure \ref{fig-ROM_GMs_NOvi} shows the column densities of \ion{O}{6} as a function of radius for our 39 R25 MW-mass galaxies. Red and grey solid lines represent the column densities of quenched and star forming galaxies within the R25 sample, respectively. The COS-Halos dataset is plotted on top in black, with red squares and blue circles distinguishing between elliptical and spiral galaxies. Upper limits are designated with arrows and unfilled markers. The column densities of \ion{o}{6} for the four zoom-in GM galaxies are shown in solid black lines. 

Figure \ref{fig-ROM_GMs_NOvi} shows that our simulations reliably reproduce the column densities of \ion{O}{6} in the CGM. While this agreement is a substantial improvement over previous simulations, which significantly underpredict NOVI, we stress that the lack of high temperature metal-line cooling in our simulations could be artificially boosting \ion{O}{6} abundances. However, given that the inclusion of high temperature metal cooling would only decrease the cooling time of gas by a factor of $\sim$ 2 at temperatures and metallicities relevant to this study \citep{Shen2010}, it is unlikely to be the dominant effect in setting N$_{\rm OVI}$, particularly when compared to AGN and halo mass, which we demonstrate can change NOVI by factors of 10 or more. We see this in both the R25 galaxies, which in addition to providing evidence for this initial result also gives cosmological credence to our suite of GM galaxies, and our four GM galaxies that include BH physics. Most significantly, we note that \textit{the column densities of \ion{O}{6} in the CGM of these galaxies does not depend on the assembly history of the galaxy.}  All of our galaxies well match the \ion{O}{6} observations despite their differing assembly histories.

Following Figure 10 of \cite{Oppenheimer2016}, Figure \ref{fig-oppenheimer} shows the average ionization fractions for all the ionization states of oxygen within three mass ranges: low mass (5 $\times$ 10$^{10}$ \textemdash 5 $\times$ 10$^{11}$ M$_{\odot}$), Milky Way-mass (5 $\times$ 10$^{11}$ \textemdash 2 $\times$ 10$^{12}$ M$_{\odot}$), and high mass (2 $\times$ 10$^{12}$ \textemdash 2 $\times$ 10$^{13}$ M$_{\odot}$). These three mass ranges include galaxies in R25 outside our sample of 39 COS-Halo mass galaxies. Dark purple, light purple, red, orange, yellow, green, light blue, dark blue, and gray indicate the oxygen ions, \ion{O}{1}, \ion{O}{2}, \ion{O}{3}, \ion{O}{4}, \ion{O}{5}, \ion{O}{6}, \ion{O}{7}, \ion{O}{8}, and \ion{O}{9}, respectively. The average ion fraction for each ion of oxygen is shown to the right of each column for the designated mass bin in its corresponding color. Ion fractions are in order from the top, highest to lowest. From the figure, we see that the \ion{O}{6} fraction (in green) decreases from the MW-mass range to the high mass regime due to the increase in virial temperature of higher mass galaxies, which moves from a value close to the ionization peak for \ion{O}{6}, T $\sim$ 10$^{5.5} K$, to 10$^{6.3}$ K. Similarly, Figure \ref{fig-highmass_Novi} shows the column densities of \ion{O}{6} for only the highest mass galaxies in R25 (2 $\times$ 10$^{12}$ \textless M$_{halo}$ \textless 2 $\times$ 10$^{13}$ M$_{\odot}$). Lines of N$_{\rm OVI}$ are colored by halo mass, with light red being the least massive and dark red denoting the highest mass galaxies. COS-Halos observations are plotted on top as in Figure \ref{fig-ROM_GMs_NOvi}. Figure \ref{fig-highmass_Novi} confirms that as galaxy virial mass increases, column densities of \ion{O}{6} decrease. This finding is consistent with the results of \cite{Oppenheimer2016} which determined that \ion{O}{6} acts as a tracer for the virial temperature of a galaxy. From this study, we determine that the star formation properties of the galaxy do not correlate with the evolution of \ion{O}{6} in the CGM. This result does not include local photoionization effects from the galaxy's star formation or AGN on the CGM. When these effects are included, especially the effect of `AGN flickering,’ \cite{Oppenheimer2018} finds a significant increase in CGM OVI column density. Instead, it appears that the mass of the galaxy, as it affects its virial temperature (Table \ref{table:BHdata}), plays a more significant role in determining the column density of \ion{O}{6} seen in the CGM of the R25 galaxies.

\subsection{Metal Transport by the SMBH}
\label{Result:metalsbyBH}

With both the R25 galaxies and zoom-in GMs, we have been able to examine the effects of star formation on the CGM. However, the zoom-in galaxies additionally offer us a controlled environment with which to more directly probe the impact of BH physics on the CGM. We examine the column densities of \ion{O}{6} in the CGM in our 4 zoom-in galaxies \textit{without} BH physics and compare them to the cases where BH physics is included. Figure \ref{fig-GMs_NOvi} shows the column densities of \ion{O}{6} in the CGM of all four of our zoom-in galaxies with BH physics (solid lines) and without (dashed lines). P0 and GM1 are light and dark blue, respectively, with GM2 and GM3 in dark and light red, as before. We can see that in the cases where BH physics is not included (dashed lines), the values of N$_{\rm OVI}$ are significantly lower implying that the presence of the SMBH must play an important role in populating \ion{O}{6} in the CGM. We look to the temperature, oxygen mass, density, and metallicity of the CGM to investigate the cause of this decrease in \ion{O}{6}.

Figure \ref{fig-GMs_profiles} shows the temperature (\textit{Upper Left}), density (\textit{Upper Right}), total mass in oxygen (\textit{Bottom Left}), and metallicity (\textit{Bottom Right}) profiles of CGM in our 4 GMs with and without BH physics (colors and line styles as in Figure \ref{fig-GMs_NOvi}). From the upper plots in Figure \ref{fig-GMs_profiles}, we see that the temperatures and densities of the CGM in our GM galaxies are not significantly changed by the lack of a SMBH. However, as we examine the bottom panels, we note a distinct difference. The CGM of the galaxies without BH physics have significantly less oxygen mass and are lower in metallicity. It appears that rather than energetically changing the temperature or physical modifying the gas density in the CGM, the lack of BH physics in these galaxies results in CGM with significant lack of metals. We look to the disk of the galaxy for more clues about this difference. Figure \ref{fig-disk_Z} (colors and line styles as in Figure \ref{fig-GMs_NOvi}) shows that, in the galaxies without BH physics (dashed lines), there is a large reservoir of metals being created near the center of the disk that is not being propogated outwards. It is the lack of SMBH feedback in these galaxies that is resulting in CGM that are severely lacking in metals. 

Figure \ref{phasediagrams} shows the phase diagrams of the CGM of the 4 zoom-in galaxies both with (\textit{Left Column}) and without BH physics (\textit{Right Column}). Examining the CGM phase diagrams for the GMs that \text{include} BH physics, we note the following key differences. First, there is decreasing overall mass from the uppermost (P0) to lowermost (GM3) figure. We can attribute this difference to the slight decrease in total halo mass from P0 to GM3 (Table \ref{table:BHdata}) and to the fact that both GM2 and GM3 are quenched galaxies. 

Second, the amount of cool, dense gas (T \textless 10$^{4.5}$, n$_H$ \textgreater 10$^{-3}$) in each galaxies' CGM varies. We attribute this to various characteristics of each simulation. In particular, for P0 and GM1 with BH physics much of this gas comes from some disk gas present at our definition of the CGM boundary, $R$ = 10 kpc. For GM2 with BH physics, this gas comes primarily from incoming satellite galaxies. We attribute the same reasoning to the 4 galaxies without BH physics which also have a similar structure in their CGM phase diagrams (as we explore below).

Finally, there is a significant lack of hot, dense gas (T \textgreater 10$^{5.5}$, n$_H$ \textgreater 10$^{-3}$) in the phase diagrams of GM2 and GM3, our quenched galaxies. To study this final difference, we explore the CGM phase diagrams that \textit{exclude} BH physics (\textit{Right Column} of Figure \ref{phasediagrams}). We note that the overall shapes of these phase diagrams are somewhat similar to the star forming galaxies \textit{with} BH physics. All four of these galaxies remain star forming throughout their evolution (Figure \ref{fig-sfh}b). The similarities end there, however, as the merger histories of these galaxies are characterized by a late-$z$ merger which occurs at slightly varying times for the 4 galaxies without BH physics. This late-$z$ merger is separate from the modified satellite which is still present at $z$ = 0 in each galaxy's halo.

P0 has its last significant merger (q $\sim$ 10, where q = M$_{halo}$/M$_{sat}$) at $z$ $\sim$ 0.7. GM1 has a similar minor satellite merger at $z$ $\sim$ 0.5  which increases the amount of metal in the CGM (up to $\sim$ 2 $\%$ compared to P0), but by $z$ = 0.17, the satellite galaxy has merged fully with the galaxy of the main halo. Only 0.1 $\%$ of the highest metallicity gas remains outside of 20 kpc from the galaxy, or about 10$^6$ M$_{\odot}$.  In GM2, the minor satellite galaxy merger occurs at $z$ $\sim$ 0.17 causing a large swell in the amount of metal enrichment seen in the CGM. This high metallicity gas (M$_{Z > 0.8 Z_{\odot}, R > 20 kpc}$ = 2.3 $\times$ 10$^{9}$ M$_{\odot}$) accounts for 3 $\%$ of the total CGM gas mass, the majority of which is outside of 20 kpc from the main halo's disk (still concentrated in the region of the satellite galaxy). This satellite in GM3 doesn't fully merge with the main halo until almost $z$ $\sim$ 0. We note that similar, late-$z$ mergers are present in the zoom-in galaxies with BH physics. However, their effect is less significant due to the metal enrichment caused by the SMBH.

There is a lack of the hot, dense gas in the quenched galaxies. However, we do see the hot, dense gas feature in the CGM phase diagrams of the galaxies \textit{without} BH physics, which all result in star forming, disked galaxies. Figure \ref{figure:phasediagrams_z_R} examines this difference with the same CGM phase diagrams of P0 and GM3 weighted by oxygen mass, metallicity, and distance from the center of the galaxy, with (\textit{Two Upper Rows}) and without (\textit{Two Lower Rows}) BH physics. The hot, dense gas in P0 with BH physics (\textit{Upper Row}) appears to be mostly comprised of high metallicity gas that is close to the disk (R \textless 50 kpc). Quantifying properties of this gas, we find that 3 $\%$ of the CGM gas has metallicity $Z$ \textgreater 0.8 Z$_{\odot}$ at $z$ = 0.17. Furthermore, of this 3 $\%$, nearly 30 $\%$ is farther than 20 kpc from the center of the galaxies. For GM1, the CGM is comprised of 6.7 $\%$ gas with $Z$ \textgreater 0.8 Z$_{\odot}$ with 55 $\%$ of that gas farther than 20 kpc. Contrastingly, a negligible amount of the CGM of both GM2 and GM3 have $Z$ \textgreater 0.8 Z$_{\odot}$ at $z$ = 0.17. The CGM of the four galaxies without BH physics also have small amounts of gas with $Z$ \textgreater Z$_{\odot}$, from 0.2 $\%$ in P0noBH to 0.1 $\%$ in GM3noBH, when discounting the contribution from the satellite merger at $z$ $\sim$ 0.2. These percentages of high metallicity gases in P0 and GM1 with BH physics point to metal exchange in the galaxy that is strongly dependent on the SMBH. This result is consistent with our discussion of Figure \ref{fig-disk_Z} and with \cite{Nelson2018} who also find that metal mass ejection due to the BHs in their simulations is key to their results (See Section \ref{sec-discuss} for more details).

The lack of high metallicity gas in the CGM phase diagrams of the galaxies with no BH physics (Figure \ref{figure:phasediagrams_z_R}, \textit{Right Column}) implies that metals are not being driven out of the disk. We find that feedback does not play a significant role in directly heating or excavating the CGM gas. Instead the SMBH's feedback is pivotal in \textit{transporting the metals} from the center of the galaxy out into the CGM. 

% ########## END OF RESULTS ###########
% #####################################

% #####################################
% ########### DISCUSSION ##############
% #####################################
\section{Discussion} 
\label{sec-discuss}

Our results are broadly consistent with those of \cite{Oppenheimer2016} who use a suite of EAGLE simulated galaxies to examine the bimodality of \ion{O}{6} column densities in star forming and quenched galaxies discovered by \cite{Tumlinson2011}. They argue that the star forming galaxies with M$_{halo}$ = 10$^{11}$ - 10$^{12}$ M$_{\odot}$ are most likely to exhibit high fractions of \ion{O}{6} because they have a virial temperature, T $\sim$ 10$^{5.5}$, which corresponds to the maximum \ion{O}{6} ionization fraction in collisional ionization equilibrium. Meanwhile, their quenched galaxies (M$_{halo}$ = 10$^{12}$ - 10$^{13}$ M$_{\odot}$) have high enough virial temperatures such that the dominant ionization state of oxygen is not \ion{O}{6} but rather \ion{O}{7} or above. \cite{Oppenheimer2016} argues that the \ion{O}{6} content is not a tracer of star formation directly, but rather a more direct thermometer for the temperature of the halo.

We note that the quenched galaxies in our sample have slightly smaller M$_{halo}$ than our star forming galaxies, unlike those in Oppenheimer. This difference explains the lack of bimodality in our sample. While all 4 of our zoom-in galaxies with BH physics have virial temperatures which maximize \ion{O}{6}, we looked at a sample of R25 galaxies that spanned a mass range extending to M$_{halo}$ = 2 $\times$ 10$^{13}$ M$_{\odot}$ to test the \cite{Oppenheimer2016} bimodality argument. 

Figure \ref{fig-highmass_Novi} directly shows that the column densities of \ion{O}{6} in the R25 sample indeed act as thermometer for the temperature of the halo. Furthermore, in Figure \ref{fig-oppenheimer}  we show that as the virial temperature increases in the R25 sample, oxygen is likely to be ionized to a higher ionization state than \ion{O}{6}. Examining galaxies within low, MW-, and high mass bins from the R25 suite, we see that the column densities of \ion{O}{6} decrease as the temperature which maximizes \ion{O}{6} (T = 10$^{5.5}$) is surpassed by the virial temperatures of these halos. 

This lack of bimodality contrasts with the findings of \cite{Suresh2017} and \cite{Nelson2018}. \cite{Suresh2017} examined a sample of star forming and quenched galaxies from the moving mesh-based Illustris simulation \cite{Vogelsberger2014}. The column densities of \ion{O}{6} in these galaxies reproduce the bimodality seen in \cite{Tumlinson2011}, wherein star forming galaxies have higher column density of \ion{O}{6} than quenched galaxies of the same mass. However, they find the total column densities of \ion{O}{6} are lower than expected based on the COS-Halos observations. \cite{Suresh2017} argue that the bimodality arises due to the effect of AGN feedback in their model rather than \ion{O}{6} acting as a temperature gauge for the halo virial temperature. To arrive at this result, \cite{Suresh2017} ran smaller simulation volumes which did not use their AGN prescription. In these smaller volumes, the bimodality disappeared. 

In comparison, \cite{Nelson2018} uses IllustrisTNG \citep{Marinacci2016,Naiman2017} to examine the \ion{O}{6} bimodality. This updated version of Illustris uses a new ``multi mode'' BH feedback model which allows for a thermal ``quasar'' mode at high accretion rates and a kinetic ``wind'' mode at low accretion rates. With this new AGN accretion model, the column densities of \ion{O}{6} in their galaxies match the COS-Halos observations and show the same bimodality as \cite{Tumlinson2011} and \cite{Suresh2017}. Specifically, \cite{Nelson2018} finds that there is likely more \ion{O}{6} in the CGM of galaxies if their galaxy has any of the following characteristics: higher gas fraction, higher sSFR, higher gas metallicity, bluer color, or a less massive BH. In addition, they conclude that the energy injected by their AGN in the kinetic feedback mode (low accretion rate) can significantly affect the \ion{O}{6} content of the CGM. They also conclude that BH feedback in this mode directly affects the \ion{O}{6} and results in higher \ion{O}{6} columns in star forming galaxies. They attribute this affect to the ejection of metal mass from the central galaxy and (to a lesser extent) the heating of CGM gas by energy infusion from the SMBH. 

Despite differences in their methods, both studies attribute the existence of a bimodality in the \ion{O}{6} column densities to the SMBH feedback in their simulations. We see no such effect. Our 4 zoom-in GM galaxies all have very similar characteristics (Table \ref{table:BHdata} and Figure \ref{fig-bh}) and we do not see significant differences between their column densities of \ion{O}{6}. Our results are consistent with those of \cite{Nelson2018} in that the SMBH is responsible for enriching the CGM by physically driving metals out of the disk.  

In our study, we establish that the SMBHs at the center of our galaxies are crucial for ejecting metal-enriched material out into the CGM, thereby elevating the column densities of \ion{O}{6}. This result implies that galaxies with lower mass BHs\textemdash and therefore less BH feedback\textemdash are likely to have lower metallicity gas in their CGM. In contrast, galaxies with higher mass BHs will have more metal-enriched CGM material. We may infer that varying BH properties result in the large distribution of CGM metallicities measured by observers \citep{Lehner2013,Wotta2016,Prochaska2017}. The lower right panel of Figure \ref{fig-GMs_profiles} shows that within our 4 zoom-in galaxies, we span a range of metallicities from -1.25 to solar, nearly the full range seen in observational studies. 

We predict that the early growth of the BH's mass (or more specifically, its accretion history) correlates directly with CGM metallicity. The lower panel of Figure \ref{fig-bh} shows the accretion history of the SMBHs in our 4 zoom-ins. While the accretion histories are similar up to z $\sim$ 1, they have significant differences at later times. This result is consistent with the idea that the CGM metal budget is built up at early times through BH feedback, while later BH feedback does not significantly change the amount of \ion{O}{6} in the CGM of their host galaxies. Using \textit{HST}/COS observations, \cite{Berg2018} (COS-AGN) examines the kinematics of cool gas in the CGM of both AGN and non-AGN host galaxies. They find no signature of recent AGN activity in the inner ($\lesssim$ 160 kpc) CGM of their sample, but do find kinematic differences at high impact parameters. They interpret this difference as an indicator that the CGM is built up by activity in the host galaxy at early times.

While many studies, both theoretical and observational, have sought to connect galaxy star formation rates, ISM content, and environments to CGM properties, there has been no observational study to explore a direct link between SMBH properties and the CGM. Future observations of the CGM in galaxies with well-known SMBH masses could attempt to address this missing link.

% ########## END OF DISCUSSION ###########
% ########################################

% #####################################
% ########### CONCLUSION ##############
% #####################################
\section{Summary and Conclusion}
\label{sec-conclude}

We have examined the effects of SMBH feedback and star formation history on the column densities of \ion{O}{6} in the CGM of galaxies with stellar masses between 3 $\times$ 10$^{9}$ \textemdash 3 $\times$ 10$^{11}$ M$_{\odot}$. To do so, we have used the cosmological volume {\sc Romulus25} and a zoom-in galaxy with 3 genetic modifications run with and without BH physics.

In our simulations, we determine that the SMBH transports metals into the CGM. Previous studies have examined the effect of AGN heating on the CGM as a way to raise ambient gas to a temperature that optimizes the production of \ion{O}{6} \citep{Suresh2017,McQuinn2017,Mathews2017}. Others have proposed that SMBH feedback may physically push outflows of gas from the galaxy, resulting in a higher mass CGM and therefore higher column densities of \ion{O}{6}. Neither of these cases is what we see in our simulations. \textit{Instead, our SMBH feedback propagates metal mass (but not total gas mass) into the outer halo. Furthermore, we find that \ion{O}{6} column densities depend on the virial temperature of the galaxy halo}. Relatedly, we determine that the presence of a SMBH alone cannot quench a galaxy. Rather a SMBH and additional factors, such as the presence of a satellite galaxy and/or previous mergers, are necessary for a galaxy to quench. 

The combined results of our large R25 cosmological simulation and our zoom-in galaxies with BH physics imply a mechanism by which column densities of \ion{O}{6} are set primarily by the virial temperature of the host galaxies and accretion history of the SMBH. However, we do not include a photo-ionization prescription in our simulations, which may have a small effect on the \ion{O}{6} content close to the disk of the galaxy. Furthermore, we find that \ion{O}{6} column densities in the CGM of our galaxies are not significantly affected by the evolution of the stellar disk. Their phase diagrams also show significant differences in response to their overall assembly histories, showing more overall and higher metallicity gas in the star forming cases. Despite these gas phase differences, the column densities of \ion{O}{6} remain unchanged. We conclude that the physical conditions that give rise to widespread \ion{O}{6} absorption in the CGM are not set by whether a galaxy quenches, but instead are driven by early SMBH feedback and the virial temperature of the galaxy halo.

% ####### END OF CONCLUSIONS ##########
% #####################################

\section{Acknowledgements}
   
The observations used in this work come from the COS-Halos survey, which was carried out under two HST programs: 11598 and 13033, through which support was provided by NASA through a grant from the Space Telescope Science Institute, which is operated by the Association of Universities for Research in Astronomy, Inc., under NASA contract NAS 5-26555. The simulations in this study were run on Blue Waters (under NSF PRAC award OCI-114435) and NAS. NNS and JW gratefully acknowledge helpful conversations with the following individuals: Ben Oppenheimer, Dylan Nelson, Molly Peeples, Todd Tripp, Matthew McQuinn, Alyson Brooks, Ferah Munshi, Jillian Bellovary, and Cameron Hummels. JW acknowledges partial support from a 2018 Alfred P. Sloan Research Fellowship. The analysis in this paper was priamrily done using the publicly available pynbody \citep{pynbody}  and TANGOS \citep{TANGOS}.
{{\it Facilities:} \facility{HST: COS} }

\bibliography{./SanchezCGM2018.bib}
% ADD THIS LATER TO BIB
% @misc{pynbody,
%   author = {{Pontzen}, A. and {Ro{\v s}kar}, R. and {Stinson}, G.~S. and {Woods},
%      R. and {Reed}, D.~M. and {Coles}, J. and {Quinn}, T.~R.},
%   title = "{pynbody: Astrophysics Simulation Analysis for Python}",
%   note = {Astrophysics Source Code Library, ascl:1305.002},
%   year = 2013
% }
%
% @BOOK{draine,
%    author = {{Draine}, B.~T.},
%     title = "{Physics of the Interstellar and Intergalactic Medium}",
% booktitle = {Physics of the Interstellar and Intergalactic Medium by Bruce T.~Draine.~Princeton University Press, 2011.~ISBN: 978-0-691-12214-4},
%      year = 2011,
%    adsurl = {http://adsabs.harvard.edu/abs/2011piim.book.....D},
%   adsnote = {Provided by the SAO/NASA Astrophysics Data System}
% }
%

\end{document}